\documentclass[a4paper,11pt]{article}
\pdfoutput=1 

\usepackage{array}
\usepackage{amsmath}
\usepackage{amssymb}
\usepackage{amsfonts}
\usepackage{makecell}
\usepackage{graphicx}
\usepackage{authblk}		 
\usepackage{wrapfig}
\usepackage{sidecap}
\usepackage{floatrow}		 
\usepackage{pgfplots}
\usepackage{subcaption}
\pgfplotsset{compat=1.10}	 
\usepackage[export]{adjustbox} 	
\usepackage[T1]{fontenc} 		

\newcommand{\D}{\text{d}}
\newcommand{\hc}{\text{H}}

\graphicspath{{./plots/}}

\usepackage{usetools}
\usepackage{tikzconf}
\usepackage{jheppub} 

\title{\boldmath Phenomenology of combined resummation for Higgs and Drell--Yan}

\author[]{Tanjona R. Rabemananjara}
\emailAdd{tanjona.rabemananjara@mi.infn.it}

\affiliation[]{Tif Lab, Dipartimento di Fisica, Universit\`a di Milano, and}
\affiliation[]{INFN, Sezione di Milano, Via Celoria 16, I-20133 Milano, Italy}

\abstract{We study the phenomenological impact of a recently suggested formalism for
the combination of threshold and a so-called \emph{threshold-improved} transverse momentum
resummation, by using it to improve the fixed-order results. This formalism allows 
for a systematic improvement of the transverse momentum resummation that is valid 
in the entire range of $p_T$ by the inclusion of the threshold contribution. We use 
the Borel method as a suitable prescription for defining the inverse Mellin and Fourier
transforms in the context of combined resummed expression. The study is applied
to two QCD processes, namely the Higgs boson produced via gluon fusion and
$Z$ boson production via the Drell--Yan mechanism. We compare our results to
the standard transverse momentum resummation, as well as to the fixed-order results. 
We find that the threshold-improved transverse momentum resummation leads to faster
perturbative convergence at small $p_T$ while the inclusion of threshold resummation
improves the agreement with fixed-order calculations at medium and large $p_T$. These 
effects are more pronounced in the case of Higgs which is known to have slower 
perturbative convergence.}

\begin{document}
\vspace*{1.50cm}
\maketitle
\flushbottom
\vspace*{-0.3cm}

\section{Introduction} \label{sec:introduction}

The current state-of-the-art accuracy in resummed calculations of transverse momentum 
spectra is N$^3$LL matched with fixed-order predictions (NLO/NNLO), and has been achieved 
recently for a variety of processes including Higgs boson production~\cite{Bizon:2017rah, 
Bizon:2018foh, Chen:2018pzu}, and Drell--Yan (DY)~\cite{Bizon:2019zgf, 
Ebert:2020dfc, Becher:2019bnm, Scimemi:2019cmh, Kallweit:2020gva, Wiesemann:2020gbm}. 
Since large logarithmic corrections present in transverse momentum 
and threshold resummation both originate from the emission of soft gluons, there 
have been attempts to construct a \emph{joint} formalism that simultaneously resum
logarithmic contributions that are enhanced at small $p_T$ and at partonic 
threshold~\cite{Li:1998is,Laenen:2000ij}. Such a joint resummation has been successful 
in producing phenomenological results at NLL accuracy for various processes including 
Higgs~\cite{Kulesza:2003wn} and vector boson production via DY 
mechanism~\cite{Kulesza:2002rh} which has been recently extended to 
NNLL~\cite{Marzani:2016smx}. While the aforementioned resummations were done in
Fourier-Mellin space, joint resummation in direct space has been achieved up to NNLL
accuracy using SCET~\cite{Lustermans:2016nvk}. In these references, however, the 
inclusion of soft contributions were not fully complete as will be discussed in this paper.

Recently, a combination of threshold and transverse momentum resummation has been
proposed in Ref.~\cite{Muselli:2017bad} that has the following features: it
reproduces the correct behaviour to any desired logarithmic order in the limit
$p_T \to 0$ for fixed $x$; it reduces to threshold resummation in the soft limit
$x \to 1$ for fixed $p_T$ up to power correction in $(1-x)$; and it reproduces
the total cross section at any given logarithmic order in the threshold limit
upon integration over the transverse momentum. This combined resummation relies
mainly on two ingredients:  (i) a modified transverse momentum resummation--
that henceforth we call \emph{threshold-improved} transverse momentum
resummation--which leads to threshold resummed expression upon integration over $p_T$, 
(ii) the combination of the threshold-improved $p_T$ expression with the pure 
threshold expression that takes into account all the logarithmic enhanced terms in the 
soft limit for finite $p_T$. Such a combined resummation is expected to allow for a 
systematic improvements of the transverse momentum that is valid for the entire range 
of $p_T$ but the phenomenological implications have never been studied. 

Therefore, the main goal of this paper is to separately assess the effects of the modified
transverse momentum and combined resummation on transverse momentum distributions. In order
to do this, we have to deal with issues that arise from the particular construction 
of the threshold-improved transverse momentum resummation. The first concerns the issue of 
performing the inverse Fourier-Mellin transform. In most resummation formulae~\cite{Parisi:1979se,
Curci:1979bg, Dokshitzer:1978hw, Collins:1981va, Sterman:1986aj, Collins:1984kg,
Collins:1989gx, Catani:1989ne, Catani:2000vq, Albino:2000cp, Laenen:2004pm,
deFlorian:2005fzc, Bozzi:2005wk, deFlorian:2011xf}, transverse momentum
resummation is performed in Fourier-Mellin ($N$-$b$) space where closed expression
can be found. However, in addition to the problem of Landau pole that prevents
the existence of an inverse Mellin in standard $p_T$ resummation, threshold-improved $p_T$
resummation presents additional singularities due to the interplay between the Mellin
moment $N$ and the impact parameter $b$ in the argument of the logarithms. We
show that this issue can be addressed by slightly modifying the Borel
prescription that was recently studied in Ref.~\cite{Bonvini:2008ei, Bonvini:2010tp} 
in the context of threshold resummation for DY cross sections. The second has to do
with logarithmic counting in which perturbative evolutions have to be treated in
a different way to properly account for the threshold behaviours. Finally, in
order to obtain valid predictions, we need to match the resummed expressions to the
fixed-order results. This requires the computation of the inverse Fourier-Mellin
transform order by order in the running of the coupling constant $\alpha_s$.

This paper is organized as follows. We begin in Section~\ref{sec:analytic-formulation} 
with a review of the combined threshold and threshold-improved $p_T$ resummation. 
We then describe the analytical procedures to address the issues introduced previously. 
Our phenomenological results are presented in Section~\ref{sec:pheno} in which we separately 
study the impact of the modified $p_T$ and combined resummation to the Higgs boson production 
at LHC via gluon-gluon fusion and a DY process at Tevatron. Finally, conclusions are drawn in 
Section~\ref{sec:conclusion}.

\section{Analytic formulation of combined resummation}
\label{sec:analytic-formulation}

In this section, we first provide a brief review of the combined resummation 
introduced in Ref.~\cite{Muselli:2017bad}. We then address specific issues that 
arise in the phenomenological studies of the modified $p_T$ resummation: (i) the 
need for a Fourier-Mellin inversion procedure, (ii) the treatment of the PDF evolution, 
and (iii) the matching to the fixed-order calculations.

Regardless of the fact that in this paper we are particularly interested in the
phenomenological implication of the combined resummation to the case of Higgs
and DY production, the expressions that follow are kept as general as
possible such that they apply to general colour singlet hadronic production.

\subsection{Combined transverse momentum and threshold resummation}
\label{subsec:combined-resummation}

Consider the inclusive hard-scattering process $h_1 + h_2 \to F(Q, p_T)+X$,
where the collision of two hadrons $h_1$ and $h_2$ produces a final-state system
$F$ with an invariant mass $Q$ and a transverse momentum $p_T$ accompanied by
an arbitrary final state $X$. As mentioned earlier, in this paper we study the
case where the final state $F$ can be either a Higgs boson or a DY
lepton pair via electroweak boson production. We denote by $\sqrt{s}$ and
$\sqrt{\hat{s}}$ the center of mass-energy of the colliding hadrons and partons
respectively.

\begin{figure}[!htbp]
\captionsetup[subfigure]{aboveskip=-1.5pt,belowskip=-1.5pt} 
\centering
\begin{subfigure}{0.475\linewidth}
\includegraphics[width=\linewidth]{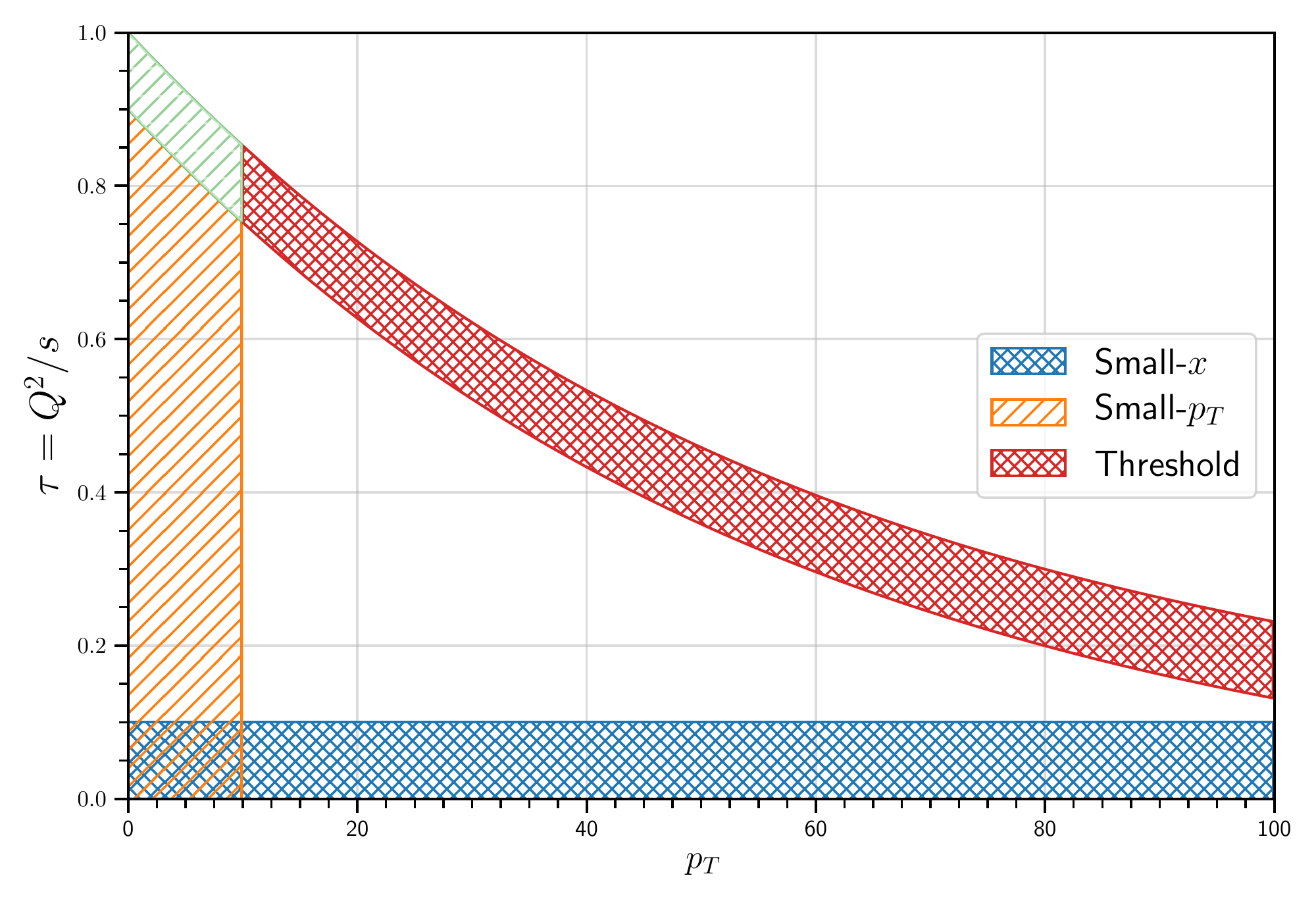} 
\end{subfigure} 
\hfil
\begin{subfigure}{0.475\linewidth}
\includegraphics[width=\linewidth]{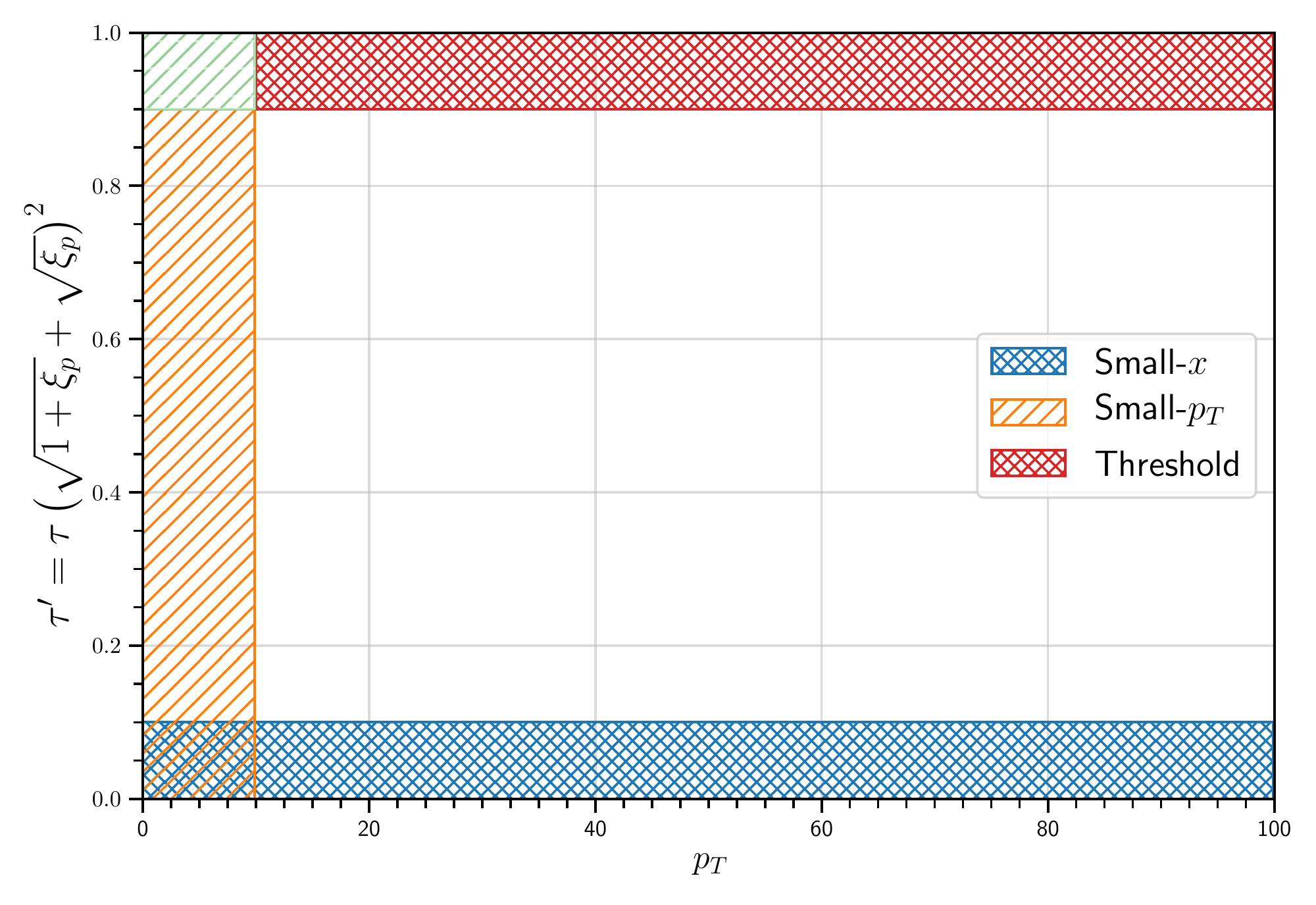} 
\end{subfigure}
\caption{Available phase space for the production of a final-state system $F$ with 
an invariant mass $Q$ and a transverse momentum $p_T$ in the three kinematic limits 
(threshold, collinear, high-energy), parametrized in terms of the kinematic variable 
$\tau = Q^2 / s$ (\emph{left}) and $\tau'$ (\emph{right}).} 
\label{fig:phase-space} 
\end{figure}

Transverse momentum distributions factorize into a convolution between a parton
luminosity and a partonic cross section which can be written in the following
way
\begin{equation}
\frac{1}{\tau^{'}} \frac{\mathrm{d} \sigma}{\mathrm{d}
\xi_p} (\xi_p, \alpha_s) = \sum_{a,b} \int_{\tau^{'}}^{1}
\frac{\mathrm{d}x}{x} \mathcal{L}_{ab} \left( \frac{\tau^{'}}{x} \right)
\frac{1}{x} \frac{\mathrm{d} \hat{\sigma}_{ab}}{\mathrm{d} \xi_p} (x, \xi_p,
\alpha_s), \label{eq:factorization}
\end{equation}
where
\begin{equation}
\tau^{'} = \frac{Q^2}{s} \left( \sqrt{1+\xi_p} + \sqrt{\xi_p} \right)^2,
\quad \text{and} \quad x = \frac{Q^2}{\hat{s}} \left( \sqrt{1+\xi_p} +
\sqrt{\xi_p} \right)^2 .
\end{equation}
The differential $p_T$ distribution
in \Eq{eq:factorization} has been expressed in terms of the dimensionless
variable $\xi_p=p_T^2/Q^2$ and, for simplicity, the renormalization and 
factorization scale dependencies have been omitted\footnote{In Ref.~\cite{Muselli:2017bad},
the invariant mass is denoted by $M$ (instead of $Q$) and $Q = (\sqrt{M^2 + p_T^2} + p_T)$.}. 
The sum is over the different partonic channels where $a$ and $b$ denote partons. 
As it turns out that in Mellin space, the convolution becomes a simple product, 
\Eq{eq:factorization} can be re-written as:
\begin{equation} \frac{\mathrm{d} \sigma}{\mathrm{d} \xi_p}
(N, \xi_p, \alpha_s) = \sum_{a,b} \mathcal{L}_{ab}(N)  \frac{\mathrm{d}
\hat{\sigma}_{ab}}{\mathrm{d} \xi_p} (N, \xi_p, \alpha_s), \label{eq:Mellin}
\end{equation}
given that
\begin{subequations}
\label{eq:Mellin-transform}
\begin{align} \frac{\mathrm{d} \sigma}{\mathrm{d} \xi_p} (N, \xi_p, \alpha_s) &=
\int_{0}^{1} \mathrm{d}\tau^{\prime} \: (\tau^{\prime})^{N-1} \frac{1}{\tau^{'}}
\frac{\mathrm{d} \sigma}{\mathrm{d} \xi_p} (\tau, \xi_p, \alpha_s),
\label{subeq:mellin-hadronic}\\ \frac{\mathrm{d}
\hat{\sigma}_{ab}}{\mathrm{d} \xi_p} (N, \xi_p, \alpha_s) &= \int_{0}^{1}
\mathrm{d}x \: x^{N-1} \frac{1}{x} \frac{\mathrm{d}
\hat{\sigma}_{ab}}{\mathrm{d} \xi_p} (x, \xi_p, \alpha_s).
\label{subeq:mellin-partonic}
\end{align}
\end{subequations}
Notice here that the cross section and its Mellin transform are denoted
with the same symbol and only distinguished trough their arguments. It is 
worth stressing that the Mellin transform of the hadronic cross section 
in~\Eq{subeq:mellin-hadronic} is taken w.r.t.  the scaling variable 
$\tau^{\prime}$ while the Mellin transform of the partonic one 
in~\Eq{subeq:mellin-partonic} is taken w.r.t. the partonic variable $x$. 
Thanks to this choice of variables, one can take simultaneously the Mellin 
and Fourier transform of the cross section. Indeed, $\tau^{\prime}$ and
$p_T$ are independent variables as $\tau^{\prime}$ ranges from $0 \leq
\tau^{\prime} \leq 1$ and $p_T$ ranges from $0 \leq p_T \leq \infty$. The
available phase induced by such choice of variables is depicted in the 
right-hand side of~\Fig{fig:phase-space}.

The partonic part in~\Eq{eq:Mellin} is constructed by combining the threshold-improved
transverse momentum, to be discussed below, with the threshold resummation. The two 
resummations are combined through a profile matching function which is chosen such that 
the combined result reproduces transverse momentum and threshold resummation at small $p_T$
and large $x$ respectively. One possible expression 
of $\mathrm{d} \hat{\sigma}_{ab} / \mathrm{d} \xi_p$ has been proposed 
in~Ref.~\cite{Muselli:2017bad}
\begin{equation}
\frac{\mathrm{d} \hat{\sigma}_{ab}}{\mathrm{d} \xi_p} (N,
\xi_p, \alpha_s) = \left( 1- \mathrm{T} (N, \xi_p) \right) \frac{\mathrm{d}
\hat{\sigma}_{ab}^{\text{tr'}}}{\mathrm{d} \xi_p} (N, \xi_p, \alpha_s) +
\mathrm{T} (N, \xi_p) \frac{\mathrm{d}
\hat{\sigma}_{ab}^{\text{thrs}}}{\mathrm{d} \xi_p} (N, \xi_p, \alpha_s),
\label{eq:combined_expr}
\end{equation}
where the profile matching function is defined as
\begin{equation}
\mathrm{T} (N, \xi_p) = \frac{N^k\xi_p^m}{1+N^k \xi_p^m}.
\label{eq:matching_function}
\end{equation}
The values of $k$ and $m$ can be chosen arbitrarily provided that $m < k$. This 
is because the combined resummation produces results that differ from the 
threshold-improved $p_T$ resummation by $\mathcal{O}(\xi_p^m)$ corrections 
when $\xi_p \to 0$, and from the threshold resummation by
$\mathcal{O}(1/N^k)$ corrections when $N \to \infty$. As mentioned in the original
reference~\cite{Muselli:2017bad}, combined resummation resums small-$p_T$ logarithms 
up to NNLL accuracy, while threshold ones are resummed up to NNLL* according
to the counting in Table 1 of Ref.~\cite{Bonvini:2014qga}.

The threshold part $\mathrm{d}\hat{\sigma}_{ab}^{\text{thrs}} / \mathrm{d} \xi_p$ 
in \Eq{eq:combined_expr} resums logarithms of the form 
$\alpha_s^n \ln^m (N)$ that are enhanced in the large-$N$ limit for fixed $p_T$
and is given in~refs.~\cite{Forte:2002ni,Bonvini:2012az} for Higgs and in~
refs.~\cite{Catani:1989ne, Albino:2000cp, Collins:1989gx, deFlorian:2005fzc}
for DY production. On the other hand, $\mathrm{d}\hat{\sigma}_{ab}^{\text{tr'}} / 
\mathrm{d} \xi_p$ represents the improved version of the standard transverse momentum 
resummation from which it differs by the inclusion of soft contributions that are 
power-suppressed terms in the small-$p_T$ limit for fixed $N$. It provides a modified 
transverse momentum expression that reproduces threshold resummation upon integration 
over $p_T$.

While the phenomenological study of the threshold resummation does not pose new 
problems, technical issues arise in the modified $p_T$ resummation that require 
specific care. Hence, the following sections are mainly devoted to the threshold-improved
$p_T$ resummation. First, in order to fully take into account the kinematics constraint 
on the transverse momentum conservation, it is advantageous to work in Fourier space where 
the part that resums small-$p_T$ logarithms in~\Eq{eq:combined_expr} yields
\begin{equation}
\frac{\mathrm{d} \hat{\sigma}_{ab}^{\text{tr'}}}{\mathrm{d} \xi_p} (N, \xi_p, \alpha_s)
= \sigma_0 \left( \sqrt{1+\xi_p} + \sqrt{\xi_p} \right)^{2N} \int_{0}^{\infty}
\mathrm{d} \hat{b} \, \frac{\hat{b}}{2}  J_0(\hat{b} \sqrt{\xi_p})
\Sigma_{ab} (N, \lambda_\chi(\hat{b})).  
\label{eq:consistent-resummed}
\end{equation}
In the above expression, $\sigma_0$ represents the leading order
Born cross section which is given in \cite{Bozzi:2005wk} for Higgs and
\cite{Bozzi:2010xn} for DY; $J_0$ is the Bessel function while $\hat{b} = b Q$ 
represents the modified conjugate variable to $\xi_p$ in Fourier space. The 
resummed partonic cross section is represented by $\Sigma_{ab}$ where the $\hat{b}$ 
dependence is embodied in $\lambda_\chi$ which is defined as
\begin{equation}
\lambda_\chi
(\hat{b}) = \alpha_s \beta_0 \ln(\chi), \quad \text{where} \quad \chi =
\bar{N}^2 + \frac{\hat{b}^2}{b_0^2},
\label{eq:log-b} 
\end{equation} 
with the modified Mellin variable defined as $\bar{N} = N \exp(- \gamma_E)$ and
$b_0 = 2 \exp(-\gamma_E)$. Notice that in comparison to the standard transverse 
momentum resummation, now the logarithm which is being resummed also depends on the 
Mellin variable $N$.

The logarithmically enhanced terms in \Eq{eq:consistent-resummed} are contained 
in the resummed expression $\Sigma_{ab}$ which can be organized in the following 
way: 
\begin{equation}
\Sigma_{ab} (N, \lambda_\chi) = \sum_{\left\lbrace \mathcal{P} \right\rbrace
} \mathcal{\tilde{H}}_{ab}^{\left\lbrace \mathcal{P} \right\rbrace } (N,
\chi) \exp(\mathcal{G}_{ab}^{\left\lbrace \mathcal{P} \right\rbrace } (N,
\lambda_\chi)), 
\label{eq:resummed-cons} 
\end{equation} 
where $\mathcal{P}$ denotes the different projectors of the LO anomalous dimension 
in the matrix flavour space. On one hand, the function $\mathcal{G}$ contains 
all the large logarithms coming from the Sudakov form factor, the evolution 
of the PDFs, and the evolution of the coefficient functions. Its complete 
expression is derived in Section~\ref{subsec:evolution-pdf}. On the other hand, 
the function $\mathcal{\tilde{H}}$ contains all the terms that behave as a 
constant in the large-$\hat{b}$ limit.

\subsection{Inverse Fourier-Mellin} 
\label{sec:mellin-inverse}

In order to get physical cross sections, we have to compute the inverse
Fourier-Mellin transform of the ($N$-$b$) space resummed expression. Mellin and/or 
Fourier resummed expressions generally give rise to technical complications when 
it comes to performing the inverse transform. Because the resummation corresponds to 
the asymptotic sum of a divergent series of $p_T$-space contributions~\cite{Bozzi:2005wk},
the Fourier-Mellin integral contains singularities.

In a standard procedure, Fourier and Mellin inverse transform are computed using
the Minimal prescription (MP)~\cite{Catani:1996yz}. In the case of inverse Fourier
transform, this amounts to finding a $\hat{b}$ contour that avoids the singularities; 
while for the inverse Mellin transform, this amounts to defining an integration path 
that passes to the left of the branch but to the right of all singularities as shown 
in~\Fig{fig:example-1}. This prescription defines the resummed hadronic cross 
section as 
\begin{equation}
\frac{\mathrm{d} \sigma}{\mathrm{d} \xi_p} (\xi_p, \alpha_s) =
\frac{\tau^{\prime}}{2 \pi \mathrm{i}} \sum_{a,b} \int_{\text{MP}} \mathrm{d}N \:
(\tau^{\prime})^{-N} \mathcal{L}_{ab}(N)  \frac{\mathrm{d}
\hat{\sigma}_{ab}}{\mathrm{d} \xi_p} (N, \xi_p, \alpha_s),
\label{eq:inverse-mellin} 
\end{equation} 
As a result, cross sections obtained through such a procedure is finite. However, 
in addition to the branch cut related to the Landau pole, the integrand 
in~\Eq{eq:consistent-resummed} contains additional singularities when
\begin{equation}
\mathfrak{Re} \left( \bar{N}^2 + \frac{\hat{b}^2}{b_0^2} \right) \le 0,
\qquad \text{and} \qquad \mathfrak{Im} \left( \bar{N}^2 +
\frac{\hat{b}^2}{b_0^2} \right) = 0.  
\label{eq:extra-singularities}
\end{equation}
This means that the location of the $N$-space singularities depends on $\hat{b}$.
Therefore, it is non-trivial to find a contour deformation that leaves all
the singularities to the left and the branch cut to the right.

An alternative approach consists on summing the divergent series using the Borel
method which was first introduced in Refs.~\cite{Catani:1996yz, Forte:2006mi,
Abbate:2007qv}. Recently, it was shown in Ref.~\cite{Bonvini:2008ei} in the 
context of resummation of transverse momentum distributions that the 
inverse Fourier-Mellin can be performed by first resumming the divergent series
using the Borel method and then removing the divergence in the inverse Borel
transform. This permits the computation of the asymptotic result the divergent
series is tending to. Such a procedure can be slightly modified to work with
transverse momentum resummation~\cite{Muselli:2017phd}. The steps are as follows.
First, we expand the resummed component $\Sigma_{ab}(N, \chi, \alpha_s)$ as a 
series in $\bar{\alpha}_s \ln (\chi)$ (with $\bar{\alpha}_s = \alpha_s \beta_0$) 
and tackle directly the divergence using the Borel summation. This then leads to 
a $N$-space expression that can be inverted using the MP. Hence, we have
\begin{equation} 
\Sigma (N, \lambda_\chi
(\hat{b}), \alpha_\mathrm{s}) = \sum_{k=0}^{\infty} h_k(N,
\alpha_\mathrm{s}) \bar{\alpha}_\mathrm{s}^k \ln^k \chi (\hat{b}) =
\sum_{k=0}^{\infty} h_k(N, \alpha_\mathrm{s}) \bar{\alpha}_\mathrm{s}^k
\ln^k \left( \bar{N}^2 + \frac{\hat{b}^2}{b_0^2} \right) \text{.}
\label{eq:series-of-sigma} 
\end{equation} 
When $k=0$ the zeroth order coefficient $h_0 (N, \alpha_s)$ is just a constant. 
Putting back the above equation into the partonic resummed expression in 
\Eq{eq:consistent-resummed}, exchanging the integral and the sum, and finally 
using the trick  $\mathrm{d}^k \chi^\epsilon / \mathrm{d} \epsilon^k 
|_{\epsilon \to 0} = \ln^k \chi$, we get to the following result
\begin{subequations} 
\begin{align} 
\frac{\mathrm{d}
\hat{\sigma}'}{\mathrm{d} \xi_p} (N, \xi_p, \alpha_s) &=
\sum_{k=0}^\infty h_k(N, \alpha_s) \bar{\alpha}_\mathrm{s}^k \left.
\frac{\partial^k}{\partial \epsilon^k} \int\limits_0^\infty
\mathrm{d} \hat{b} \frac{\hat{b}}{2} J_0 \left( \hat{b} \sqrt{\xi_p}
\right) \left( \bar{N}^2 + \frac{\hat{b}^2}{b_0^2} \right)^\epsilon
\right|_{\epsilon \to 0}, \label{eq:series-sigma} \\ &=
\sum_{k=0}^\infty h_k(N, \alpha_s) \bar{\alpha}_\mathrm{s}^k \left.
\frac{\partial^k}{\partial \epsilon^k} \frac{1}{b_0^{2
\epsilon}} \int\limits_0^\infty \mathrm{d} \hat{b}
\frac{\hat{b}}{2} J_0 \left( \hat{b} \sqrt{\xi_p} \right) \left(
4 N^2 + \hat{b}^2 \right)^\epsilon \right|_{\epsilon \to 0} .
\label{eq:series-bhat} 
\end{align} 
\end{subequations} 
Remarks are in order concerning the above equations: $(\mathrm{i})$
$\mathrm{d} \hat{\sigma}'/\mathrm{d}\xi_p$ is just
$\mathrm{d} \hat{\sigma}^{\text{tr'}}/\mathrm{d}\xi_p$ without the
kinematic factor $\left( \sqrt{1+\xi_p} + \sqrt{\xi_p} \right)^{2N}$
and the Born level cross section $\sigma_0$, $(\mathrm{ii})$ $b_0$
has been factorized out such that the integrand in
\Eq{eq:series-sigma} is a function of $N$ instead of $\bar{N}$. This
allows us to perform the inverse Fourier transform analytically
order by order. Hence, \Eq{eq:series-bhat} becomes 
\begin{equation}
\frac{\mathrm{d} \hat{\sigma}'}{\mathrm{d} \xi_p} (N, \xi_p,
\alpha_s) = \sum_{k=0}^\infty h_k(N, \alpha_s)
\bar{\alpha}_\mathrm{s}^k \left. \frac{\partial^k}{\partial
\epsilon^k} M(N, \xi_p, \epsilon) \right|_{\epsilon \to 0},
\end{equation} 
where the function $M$ is defined as
\begin{equation} 
M(N, \xi_p, \epsilon) = 2 \, \mathrm{e}^{2
\gamma_E \epsilon} \left( \frac{N}{\sqrt{\xi_p}}
\right)^{1+\epsilon} \frac{ \mathrm{K}_{1+\epsilon} \left( 2
N \sqrt{\xi_p} \right)}{\Gamma (-\epsilon)}.
\label{eq:bhat-inverse} 
\end{equation} 
The function $\mathrm{K}$
denotes the modified Bessel function of the second kind. We can then
get rid of the derivatives and the limit by performing a contour
integral, 
\begin{equation} 
\left. \frac{\partial^k}{\partial
\epsilon^k} M(N, \xi_p, \epsilon) \right|_{\epsilon \to 0} =
\frac{k!}{2 \pi \mathrm{i}} \oint_H \frac{\mathrm{d} \xi}{\xi^{1+k}}
M( N, \xi_p, \xi ) \text{,} 
\end{equation} where $H$ represents a
contour enclosing the singularity at $\xi = 0$.

Having expanded the function $\Sigma$ allowed us to perform the inverse Fourier
transform, but it leaves us with a series representation of the result. Thinking
of the numerical limitations of series representations, we would rather trade
the sum for an integral. This basically calls for an analytic continuation,
which in this case is achieved using a Borel summation and a subsequent Borel
back-transformation. First, writing down the series as 
\begin{equation}
\frac{\mathrm{d} \hat{\sigma}'}{\mathrm{d} \xi_p} (N, \xi_p, \alpha_s)
\equiv A(\bar{\alpha}_s) = \sum_{k=0}^\infty h_k(N, \alpha_s) \frac{k!}{2
\pi \mathrm{i}} \oint_H \frac{\mathrm{d} \xi}{\xi} M( N, \xi_p, \xi ) \left(
\frac{\bar{\alpha}_s}{\xi} \right)^k \text{.} 
\end{equation} 
The corresponding Borel sum is then given by 
\begin{equation} 
\begin{split} 
\mathcal{B} \left[ A
(w) \right] &= \frac{1}{2 \pi \mathrm{i}} \sum_{k=0}^\infty h_k(N,
\alpha_\mathrm{s}) \oint_H \frac{\mathrm{d} \xi}{\xi} M( N, \xi_p, \xi )
\left( \frac{w}{\xi} \right)^k \\ &= \frac{1}{2 \pi \mathrm{i}} \oint_H
\frac{\mathrm{d} \xi}{\xi} M( N, \xi_p, \xi ) \Sigma \left( N, \frac{w}{\xi},
\alpha_\mathrm{s} \right) \text{,} 
\end{split} 
\end{equation} 
where we resummed back the expansion of $\Sigma$ using its definition in
Eq.~\eqref{eq:series-of-sigma}.  We finally perform the Borel
back-transformation, 
\begin{subequations} 
\begin{align} \frac{\mathrm{d}
\hat{\sigma}'}{\mathrm{d} \xi_p} (N, \xi_p, \alpha_s) &=
\int\limits_0^\infty \mathrm{d} w \, \mathrm{e}^{-w} \mathcal{B} \left[
A (w \bar{\alpha}_\mathrm{s}) \right] = \frac{1}{\bar{\alpha}_\mathrm{s}} 
\int\limits_0^\infty \mathrm{d} w' \, \mathrm{e}^{-w' / \bar{\alpha}_\mathrm{s}} 
\mathcal{B} \left[ A (w') \right] \label{subeq:final-result-1} \\ &=
\frac{1}{\bar{\alpha}_\mathrm{s}}  \int\limits_0^\infty \mathrm{d} w \,
\mathrm{e}^{-w / \bar{\alpha}_\mathrm{s}} \frac{1}{2 \pi \mathrm{i}} \oint_H
\frac{\mathrm{d} \xi}{\xi} M( N, \xi_p, \xi ) \Sigma \left( N,
\frac{w}{\xi}, \alpha_\mathrm{s} \right) \text{,} 
\label{eq:final-result}
\end{align} 
\end{subequations} which represents our final resummed expression
when multiplied by the factor $\sigma_0 (\sqrt{1+\xi_p} + \sqrt{\xi_p})^{2N}$
that we left out from the beginning. After cutting off the $w$-integration at
some finite number $C$, we have the final expression \label{eq:final-integral}
\begin{equation} 
\frac{\mathrm{d} \hat{\sigma}_{ab}^{\text{tr'}}}{\mathrm{d}
\xi_p} = \frac{\sigma_0}{\bar{\alpha}_\mathrm{s}} \left( \sqrt{1+\xi_p} + 
\sqrt{\xi_p} \right)^{2N} \int\limits_0^C \mathrm{d} w \,
\frac{\mathrm{e}^{-w / \bar{\alpha}_\mathrm{s}}}{2 \pi \mathrm{i}} \oint_H
\frac{\mathrm{d} \xi}{\xi} M( N, \xi_p, \xi ) \Sigma_{ab} \left( N,
\frac{w}{\xi}, \alpha_\mathrm{s} \right).  \label{eq:final-resummed}
\end{equation} 
The Borel and Minimal Prescription differ in the way
higher-twist behaviour of the resummed series is handled. However, in practice
and as will be demonstrated in the next sections, this difference is negligible
at collider energies we are interested in. The inclusion or exclusion of
subleading terms in the Borel expression of \Eq{eq:final-resummed} is regulated
by the cutoff $C$ which can be used to estimate the ambiguity of the resummation
procedure. The larger the value of $C$ gets, the more we include
power-correction terms in the series. However, as $C$ gets bigger the integral
in \Eq{eq:final-resummed} becomes unstable and varies a lot for small changes of
$C$ eventually spoiling the accuracy of the result. This is due to the fact that
the series in \Eq{eq:series-of-sigma} is not Borel summable.

The following subsections deal with the numerical implementation of the
procedure described above and, specifically, comment on the proper choice of the
cutoff $C$ regulating the truncation of the Borel integral making it convergent.

\subsubsection{Contour and Borel integration} 
\label{subsec:contour-integral}

By examining the singularity structure of the integrand in
\Eq{eq:final-resummed} (see \Fig{fig:example-2}),  we see that the contour
integral can be written as an integral over the domain $[0,1]$, 
\begin{equation}
\frac{1}{2 \pi \mathrm{i}} \oint_H \mathrm{d} \xi \, f (\xi) = R
\int\limits_0^1 \mathrm{d} \varphi \, \mathrm{e}^{2 \mathrm{i} \pi \varphi}
f ( \xi (\varphi) ) \text{,} 
\end{equation} 
where $f$ represents the integrand. The contour is parametrized by 
$\xi (\varphi) = x_0 + R \exp (2 \mathrm{i} \pi \varphi)$ which is a 
circle around the real point $\xi = x_0 \in \mathbb{R}$ with a radius 
$R > 0$, in the positive direction around $x_0$. The branch cut and poles of 
$f$ are now mapped  in terms of the variable $\xi$ into a branch cut on the 
real axis from zero to $w$(see \Fig{fig:example-2}). In practice, we choose 
$x_0 = w/2$ and $R=w/2+\zeta$ ($\zeta \in \mathbb{R}^{*+}$). The value of $\zeta$
and the cutoff $C$ must be chosen in such a way that the circle does not
intersect the branch cut in $\xi$-space. Based on the \emph{twist-4}
argument~\cite{Bonvini:2012sh}, the value of the cutoff should be chosen
according to $C \geq 2$. Throughout our implementation, we chose the minimal
value for $C$: i.e $C=2$.  
\begin{figure}[!h] 
\centering 
\begin{floatrow}
\ffigbox
{
	\begin{tikzpicture} 
	\begin{axis}[ axis lines=middle, xmin=-45,xmax=200,
	ymin=-25,ymax=25, xtick={0}, ytick={0}, xlabel={$\mathfrak{Re}(N)$}, xlabel
	style={anchor=north}, ylabel={$\mathfrak{Im}(N)$}, ylabel
	style={anchor=east}, enlarge y limits=upper, samples=2 ]
	\addplot[cblue,thick,fermionbar,domain=-25:100,name path=plot1] {20-x/5};
	\addplot[cblue,very thick] coordinates {(100,-1.5)(100,1.5)};
	\addplot[cblue,thick,fermion,domain=-25:100,name path=plot2] {-20+x/5}
	node[above]{$N_0$}; 
	\addplot[corange,zigline,thick,domain=125:195] {0};
	\addplot[corange,very thick] coordinates {(125,-1.5)(125,1.5)}
	node[below=0.25cm]{$N_L$}; 
	\end{axis} 
	\end{tikzpicture} 
}%
{
\caption{Singularity structures related to the Landau pole and the Minimal
prescription path for the Mellin inverse integration.}
\label{fig:example-1}	
} 
\hfill 
\ffigbox 
{
	\begin{tikzpicture} 
	\begin{axis}[ axis lines=middle,
	xmin=-1.5,xmax=1.5, ymin=-1.5,ymax=1.5, xtick={-1,1},
	ytick={-1,1}, xlabel={$\mathfrak{Re}(\xi)$}, xlabel
	style={anchor=north}, ylabel={$\mathfrak{Im}(\xi)$}, ylabel
	style={anchor=east}, enlarge y limits=upper, samples=2 ]
	\addplot[cblue,thick] (0.3,0) circle [radius=0.65];
	\addplot[corange,zigline,thick,domain=0:0.6] {0}
	node[above]{$w$}; 
	\addplot[thick] coordinates
	{(0.3,-0.1)(0.3,0.1)} node[below=0.25cm]{$\frac{w}{2}$}; 
	\end{axis}
	\end{tikzpicture} 
} 
{
\caption{Branch cut of the integrand term $M(N,\xi_p,\xi)
\Sigma(N,w/ \xi,\alpha_s)$ in the $w$-integral given in 
\Eq{eq:final-resummed}; $H$ is 	given by the 
blue circle.}
\label{fig:example-2}
} 
\end{floatrow} 
\end{figure}

\subsubsection{Inverse Mellin} 
\label{subsec:inverse-mellin}

Thanks to the Borel procedure, the Minimal Prescription can now be applied to
perform the Mellin back transformation. The inverse Mellin transform given by
Eq.~\eqref{eq:inverse-mellin} can be rewritten in the following way
\begingroup
\allowdisplaybreaks
\begin{subequations} 
\begin{align} 
\frac{1}{2 \pi \mathrm{i}} \int\limits_{N_0 -
\mathrm{i} \infty}^{N_0 + \mathrm{i} \infty} \mathrm{d} N \, (\tau^{\prime})^{-N}
\frac{\mathrm{d} \sigma}{\mathrm{d} \xi_p}(N) &= \frac{1}{\pi}
\operatorname{Im} \left[ \int\limits_{N_0}^{N_0 + \mathrm{i} \infty}
\mathrm{d} N \, (\tau^{\prime})^{-N} \frac{\mathrm{d} \sigma}{\mathrm{d}
\xi_p}(N) \right] \\ &= \frac{1}{\pi} \operatorname{Im} \left[ (\mathrm{i} + r)
\int\limits_0^1 \frac{\mathrm{d} u}{u} (\tau^{\prime})^{-N(u)} \frac{\mathrm{d}
\sigma}{\mathrm{d} \xi_p}( N(u) ) \right] \text{,} 
\end{align}
\end{subequations}
\endgroup
where $N_0$, as usual, must be located to the right of all
the singularities of the integrand. In the first line, we split the contour into
two pieces, the first above the real axis and the second below. Since
$\mathrm{d}\sigma/ \mathrm{d} \xi_p (\tau^{\prime*}) = [\mathrm{d}\sigma/ \mathrm{d}
\xi_p (\tau^{\prime})]^*$ is real, the second piece turns out to be the negative of
the complex conjugate of the first and hence can be combined to the former. In
the second line, we used the substitution $N(u) = N_0 + (r + \mathrm{i}) \ln u$
with $r > 0$ being an arbitrary real and positive parameter which controls the
slope of the path that enhances the numerical convergence of the integral. This
parameter must be positive, so that for large imaginary values of $N$ (i.e.\
small values of $u$), the prefactor converges to zero, 
\begin{equation}
(\tau^{\prime})^{-N(u)} = \exp ( N(u) \ln 1/\tau^{\prime} ) \text{,} 
\end{equation}
requiring the real part of $N(u)$ to be negative for $u \to 0$ (i.e.\ $N_0 - r
\ln 1/u < 0$). Notice that the following is also an equivalent transformation:
\begin{equation} \frac{1}{2 \pi \mathrm{i}} \int\limits_{N_0 - \mathrm{i}
\infty}^{N_0 + \mathrm{i} \infty} \mathrm{d} N \, (\tau^{\prime})^{-N}
\frac{\mathrm{d} \sigma}{\mathrm{d} \xi_p}(N) = \frac{1}{\pi}
\operatorname{Im} \left[ (r - \mathrm{i}) \int\limits_0^1 \frac{\mathrm{d}
u}{u} (\tau^{\prime})^{-N(u)} \frac{\mathrm{d} \sigma}{\mathrm{d} \xi_p}( N(u) )
\right] \text{,} 
\end{equation} where in this case $N(u) = N_0 + (r - \mathrm{i}) \ln u$.

\subsection{PDF evolution for threshold-improved $p_T$ resummation}
\label{subsec:evolution-pdf}

In order to compute hadronic cross sections, we need to combine the partonic
resummed cross section with the Parton Distribution Functions (PDFs) embodied in
the parton luminosity $\mathcal{L}$ in \meqs{eq:factorization} and (\ref{eq:Mellin}). 
Transverse momentum resummation requires the evaluation of the PDFs at a scale
which differs from the hard scale of the fixed-order calculation. The difference
in scale is therefore taken care by PDF evolution functions. In the
context of threshold-improved transverse momentum resummation, PDFs are evaluated at the 
scale $Q^2/ \chi$ where $\chi$, as mentioned previously, depends both on the Mellin 
moment $N$ and the impact parameter $b$. Such a choice of scale, as a result, entails 
a new way of counting logarithms in the evolution which is the subject of this section.

The explicit expression of the partonic cross section $\D\hat{\sigma}_{ab}^{\text{tr'}}
/ \mathrm{d} \xi_p$ writes as: 
\begin{align} 
& \frac{\D\hat{\sigma}_{ab}^{\text{tr'}}}{\D \xi_p} \left(N, \chi, \alpha_s \right) 
= \left( \sigma_0 \right)_{c \bar{c}} \bar{\hc}_c \left( \frac{\bar{N}^2}{\chi} ,
\alpha_s (Q^2) \right) C_{ci} \left(N, \alpha_s\left( \frac{Q^2}{\chi}
\right) \right) C_{\bar{c} j} \left(N, \alpha_s\left( \frac{Q^2}{\chi} \right) \right)
\nonumber \\ & \hspace*{0.6cm} U_{ia} \left( N, \alpha_s\left( \frac{Q^2}{\chi}
\right), \alpha_s(\mu_F^2) \right) U_{jb} \left( N, \alpha_s\left(
\frac{Q^2}{\chi} \right), \alpha_s(\mu_F^2) \right) \exp \left( S_c \left( N,
\chi, \mu_R^2 \right) \right), \label{partonic_evol} 
\end{align}
where here $\bar{\hc}$ represents the hard factor which behaves as a constant 
in the large-$\hat{b}$ limit, $C$ is the coefficient functions, $U$ is the 
evolution operators that evolve the PDFs from the scale $Q^2/\chi$ to $\mu_F$, 
and finally, $S_c$ is the Sudakov form factor. The subscripts define partonic 
indices where repeated indices are summed over. In the case of the Higgs production
via gluon fusion, for instance, $c=g$. Finally, $\mu_R$ and $\mu_F$ represent 
the renormalization and factorization scale respectively. From~\Eq{partonic_evol},
we can see that the modified resummed expression differs from the standard 
$p_T$ resummation in two ways. First, the argument of the logarithms in the 
Sudakov exponent includes the soft contribution. Second,  as mentioned above, the 
argument of the running of coupling, both in the coefficient functions and in the 
evolution operators, is computed at the scale $Q^2/\chi$ instead of $b_0^2/b^2$.

In order to determine the appropriate logarithmic counting, we work out in this 
section the solution of the evolution equation explicitly. Let us first start with 
the coefficient functions $C$ of~\Eq{partonic_evol}. One can set all the arguments 
of $\alpha_s$ in $C$ to be the same and take into account the difference in scales
in an evolution factor $R$ such that 
\begin{equation} 
C_{ci} \left(N, \alpha_s\left( \frac{Q^2}{\chi} \right) \right) = C_{ci} \left(N,
\alpha_s(Q^2) \right) R_{ci} \left(N, \alpha_s\left( \frac{Q^2}{\chi} \right),
\alpha_s(Q^2) \right), \label{evol_R} 
\end{equation} 
where 
\begin{equation}
R_{ci} \left(N, \alpha_s\left( \frac{Q^2}{\chi} \right), \alpha_s(Q^2)
\right) = \exp \left\lbrace - \int_{Q^2/\chi}^{Q^2} \frac{\D q^2}{q^2}
\frac{\beta \left( \alpha_s(q^2) \right)}{\alpha_s(q^2)} \left[ \frac{\D \ln
C_{ci} \left( N, \alpha_s(q^2) \right)}{\D \ln \alpha_s(q^2)} \right]
\right\rbrace .  
\label{R_expression} 
\end{equation}

Let us now turn to the PDF evolution. Here, we can evolve the PDFs from $Q^2 / \chi$
to $Q^2$ and then from $Q^2$ to $\mu_F^2$. In the case where $Q = \mu_F$, we have
$U_{ij} \left( N, \alpha_s(Q^2), \alpha_s(\mu_F^2) \right) = 1$. The expression of the PDF evolution can be written 
as~\cite{Vogt:2004ns}: 
\begin{subequations}
\begin{align} 
U_{ia} \left( N, \alpha_s\left( \frac{Q^2}{\chi} \right),
\alpha_s(Q^2) \right) =& V_{il} \left( N, \alpha_s\left(
\frac{Q^2}{\chi} \right) \right) U^{\text{(LO)}}_{lk} \left( N,
\alpha_s\left( \frac{Q^2}{\chi} \right), \alpha_s(Q^2) \right) \nonumber \\ 
& \tilde{V}_{ka} \left( N, \alpha_s(Q^2) \right) , 
\label{evolution-expression}
\end{align} 
\end{subequations} 
where $U^{\text{(LO)}}$ and $V$ respectively
represent the lowest and higher perturbative order solutions to the evolution
equation, and $\tilde{V}_{ka}$ denotes the $(k,a)$-element of the inverse matrix
$\mathbf{V}^{-1}$ in the flavour space. Henceforth, we shall use the boldface
notation to denote the representation in flavour matrix space. In the case of
singlet, the lowest perturbative order solution is derived by diagonalizing the
LO anomalous dimensions matrix $\mathbf{\gamma}^{(0)}$. Thus, following
refs.~\cite{Vogt:2004ns, Bozzi:2005wk}, we have 
\begin{equation}
U^{\text{(LO)}}_{lk} \left( N, \alpha_s\left( \frac{Q^2}{\chi} \right),
\alpha_s(Q^2) \right) = \sum_{r=\pm} e^{(r)}_{lk} (N) \exp \left\lbrace -
\frac{\lambda^{(0)}_r (N)}{\beta_0} \ln \left( \frac{\alpha_s (Q^2/
\chi)}{\alpha_s(Q^2)} \right) \right\rbrace , \label{lowest-order-evolution}
\end{equation} 
where $\lambda_\pm$ and $e^\pm$ represent the eigenvalues of the
singlet matrix $\mathbf{\gamma}^{(0)}$ and its projectors respectively whose
expressions are given by Eqs.~(2.27) and (2.28) of~\cite{Vogt:2004ns}. On the
other hand, the functions $V$ which take into account the higher order solutions
to the evolution equation can be perturbatively expanded as a series in
$\alpha_s$, 
\begin{equation} 
V_{il} (N, \mu) = \delta_{il} + \sum_{n=1}^{\infty} \alpha_s^{(n)}(\mu) 
V^{(n)}_{il} (N).
\label{high_order_solution_expansion} 
\end{equation} 
Each coefficient $V^{(i)}$ in \Eq{high_order_solution_expansion} can be computed 
iteratively following~\cite{Vogt:2004ns}. For instance, the first order coefficient
$\mathbf{V}^{(1)}$ is given by 
\begin{equation} 
\mathbf{V}^{(1)}(N) =
\sum_{i,j=\pm} \frac{1}{\lambda^{(0)}_j(N) - \lambda^{(0)}_i(N) - \beta_0}
\mathbf{e}^{(i)}(N) \left( \gamma^{(1)}(N) - \frac{\beta_1}{\beta_0}
\gamma^{(0)}(N) \right) \mathbf{e}^{(i)}(N), 
\end{equation} where we have introduced the coefficients $\beta_i$ 
of the QCD $\beta$-function.

Once again, one can take into account the difference in the argument of the
running of coupling of $V(N, \alpha_s (Q^2 / \chi))$ in \Eq{evolution-expression}
by introducing an evolution factor 
\begin{equation} 
V_{il} \left( N, \alpha_s\left( \frac{Q^2}{\chi} \right) \right) = V_{il} 
\left( N, \alpha_s (Q^2) \right) E_{il} \left(N, \alpha_s\left( \frac{Q^2}{\chi} 
\right), \alpha_s(Q^2) \right) 
\end{equation} 
where similarly to \Eq{evol_R},
\begin{equation} 
E_{il} \left(N, \alpha_s\left( \frac{Q^2}{\chi} \right),
\alpha_s(Q^2) \right) = \exp \left\lbrace - \int_{Q^2/\chi}^{Q^2}
\frac{\D q^2}{q^2} \frac{\beta \left( \alpha_s(q^2) \right)}{\alpha_s(q^2)}
\left[ \frac{\D \ln V_{il} \left( N, \alpha_s(q^2) \right)}{\D \ln
\alpha_s(q^2)} \right] \right\rbrace .  
\label{Ru_expression} 
\end{equation}

At this point, we can discuss the difference in counting between threshold-improved
and standard $p_T$ resummation. In the context of standard transverse
momentum resummation, the leading-order solution to the evolution equation
contributes with a single logarithm of $\hat{b}$. However, from the point of view of
the threshold resummation, the flavour diagonal anomalous dimensions carry an
additional contribution of the form $A^{p_T}_{g/q} \ln(\bar{N}^2)$. To see this,
let us compute the term in the exponent of the lowest-order solution to the
evolution equation given by \Eq{lowest-order-evolution}: 
\begin{equation}
\mathrm{G}_{\pm} (N, \chi, \alpha_s) \equiv - \frac{\lambda^{(0)}_\pm
(N)}{\beta_0} \ln \left( \frac{\alpha_s (Q^2/ \chi)}{\alpha_s(Q^2)} \right)
= \frac{\gamma^{(0)}_{gg/qq}}{\beta_0} \ln(1-\lambda_\chi)  + \mathcal{O}
\left( \alpha_s (Q^2) \right) .  
\end{equation} 
Taking the large-$N$ limit of the anomalous dimension leads to the following
result
\begin{equation}
\frac{\gamma^{(0)}_{gg/qq}}{\beta_0} \ln(1-\lambda_\chi) 
\xrightarrow{N \to \infty} - \frac{A^{p_T}_{g/q}}{\beta_0} \ln(\bar{N}) 
\ln(1-\lambda_\chi) = - \frac{A^{p_T}_{g/q}}{2 \bar{\alpha}_s \beta_0} \lambda_N
\ln(1-\lambda_\chi), 
\end{equation} 
where we have defined $\lambda_N = \bar{\alpha}_s \ln(\bar{N}^2)$. Therefore, 
in order to compute N$^k$LL threshold-improved $p_T$ resummation, 
the large-$N$ behaviour of the evolution has to be included up to N$^k$LO, which 
is not the case in the standard transverse momentum resummation as N$^{k-1}$LO 
is enough. That is, at NNLL, the solution of evolution is performed up to NNLO 
accuracy but with the NNLO anomalous dimension substituted by its large-$N$ 
behaviour.

Combining all the results, we can organize terms in such a way that we obtain
the form presented in \Eq{eq:resummed-cons}. The perturbative function
$\tilde{\mathcal{H}} = \sum_{n=0}^{\infty} \alpha_s^n \tilde{\mathcal{H}}^{(n)}$ is 
then given by 
\begin{subequations} 
\begin{align}
\tilde{\mathcal{H}}_{ab}^{\left\lbrace \mathcal{P} \right\rbrace } (N, \chi)
=& \bar{\hc}_c \left( \frac{\bar{N}^2}{\chi} , \alpha_s (Q^2) \right)
\tilde{C}_{cl} \left( N, \alpha_s(Q^2) \right) \tilde{C}_{\bar{c} m} \left( N,
\alpha_s(Q^2) \right) \nonumber \\ & \hspace*{2cm} e^{(r)}_{lk} (N)
\tilde{V}_{ka} \left( N, \alpha_s(Q^2) \right) e^{(p)}_{mn} (N) \tilde{V}_{nb}
\left( N, \alpha_s(Q^2) \right), 
\end{align} 
\end{subequations} 
where here $\left\lbrace \mathcal{P} \right\rbrace = \left\{ r,p \right\} = \pm$ 
and we have defined $\tilde{C}_{cl} = C_{ci} V_{il}$. Finally, the universal form
factor in the exponent of~\Eq{eq:resummed-cons} which contains all the
logarithmic enhanced terms is expressed as 
\begin{equation}
\mathcal{G}^{\left\lbrace \mathcal{P} \right\rbrace } (N, \lambda_\chi) = S_c +
\mathrm{G}_{r}  + \mathrm{G}_{p} + \ln \left( \tilde{R}_{cl} \tilde{R}_{cm}
\right), 
\end{equation} 
where the evolution factor $\tilde{R}_{cl}$ is defined as $\tilde{R}_{cl} = R_{ci} E_{il}$. 
The logarithmic expansion of $S_c$ starts at LL accuracy. While the expansion of 
$\mathrm{G}_{r/p}$ starts at NLL accuracy in the standard $p_T$ resummation, their large-$N$ 
behaviour already contribute at LL in the context of modified $p_T$ resummation 
in order to take into account for the soft behaviour. Similarly to the standard 
$p_T$ resummation, the flavour off-diagonal terms of $\tilde{R}_{cl}$ ($c \neq l$) 
starts to contribute at NLL. The terms in the flavour diagonal, instead, start to 
contribute at NLL as opposed to the standard resummation procedure where they start 
to contribute at NNLL. 

To summarize, the truncation of the threshold-improved transverse momentum resummation at 
a given logarithmic accuracy is defined in the following way:
\begin{itemize} 
\item At LL, we approximate the hard function $\tilde{\mathcal{H}}$ by its lowest
perturbative order and include $g_1$ in the Sudakov exponent. The
$\mathrm{G}_\pm$ functions are included up to LO with $\gamma^{(0)}$
replaced by its large-$N$ behaviour while the $\tilde{R}$ functions are
approximated to $1$.  
\item At NLL, we include $\tilde{\mathcal{H}}^{(1)}$ in $\tilde{\mathcal{H}}$ 
with the function $g_2$ in the Sudakov exponent. In addition, we include the
complete LO term in $\mathrm{G}_\pm$ together with the NLO term where
$\gamma^{(0)}$ and $\gamma^{(1)}$ are replaced the their large-$N$ behaviour. 
The full expression of $\tilde{R}_{gq}$ is also included while we only include 
the large-$N$ behaviour of $V^{(1)}(N)$ in $\tilde{R}_{gg}$.  
\item At NNLL, we include in $\tilde{\mathcal{H}}$ the coefficient $\tilde{\mathcal{H}}^{(2)}$ 
and the function $g_3$ in the Sudakov exponent. Similarly, we also include the 
complete NLO term in $\mathrm{G}_\pm$ together with the NNLO term where the anomalous 
dimensions are replaced by their large-$N$ behaviour. Finally, we add to $\tilde{R}_{gg}$ 
the large-$N$ behaviour of $V^{(2)}(N)$.  
\end{itemize}

\subsection{Finite order truncation of the resummed expression}
\label{subsec:finite-truncation}

In order to provide a valid prediction up to NNLL+NLO, we need to match the
resummed results to the fixed-order calculation. This matching procedure
guarantees the correct behaviour from perturbative calculations up to a specified
order and incorporates the large logarithms from resummation at higher-orders.
The matching consists on adding to the all-order results the perturbative
calculations truncated at a given order in $\alpha_s$ and subtract to the whole
the expansion of the resummed results at the same order. The matched expression
of the combined cross section is therefore given by
\begin{equation}
\frac{\mathrm{d} \hat{\sigma}_{ab}^{\text{match}}}{\mathrm{d} \xi_p} =
\frac{\mathrm{d} \hat{\sigma}_{ab}^{\text{resum}}}{\mathrm{d} \xi_p} +
\left[\frac{\mathrm{d} \hat{\sigma}_{ab}^{\text{F.O}}}{\mathrm{d} \xi_p}
\right]_{\mathcal{O}(\alpha_s^n)}
- \left[ \frac{\mathrm{d} \hat{\sigma}_{ab}^{\text{exp}}}{\mathrm{d} 
\xi_p} \right]_{\mathcal{O}(\alpha_s^n)}.
\label{eq:matching-expression} 
\end{equation} 
The subscripts $\mathcal{O} (\alpha_s^n)$ indicate the order at which both the fixed-order 
and the expanded expressions are truncated. Here, the notation used to denote the
fixed-order (F.O) accuracy and therefore the matched result refers to the accuracy of the $p_T$ 
distribution. In particular, the LO refers to the non-trivial order of the transverse momentum 
distribution whose integral is the NLO total cross section. In the sequel, LO and NLO will always 
refer to the accuracy of the $p_T$ spectrum. The expansion of $\Sigma_{ab}$ will give rise to a 
series of the form 
\begin{equation} 
\Sigma_{ab}^{\text{exp}} (N, \chi, \alpha_s) = (\sigma_0)_{c \bar{c}} 
\left\lbrace \delta_{ca} \delta_{\bar{c} b} + \sum_{n=1}^{\infty} \alpha_s^n 
\left[  \Sigma_{ab}^{(n)} (N, \chi) + \mathcal{H}^{(n)}_{ab} (N, \chi) \right]  
\right\rbrace .  
\label{eq:truncated_expr} 
\end{equation} 
The perturbative coefficients $\Sigma^{(n)}$ are polynomials in the logarithm
variable $\ln(\chi)$. They vanish when $\ln(\chi)=0$ (i.e. $\hat{b} =
-\bar{N}^2 b_0^2$) and lead to threshold resummation when $b=0$. On the other hand, 
the function $\mathcal{H}^{(n)}$ contains all the terms that behave as constants 
in the large-$\hat{b}$ limit. The explicit expressions of the two perturbative
coefficient functions $\Sigma^{(1)}$ and $\Sigma^{(2)}$ which contribute in 
\Eq{eq:truncated_expr} are given by 
\begin{subequations} 
\label{pert_expansion} 
\begin{align}
\Sigma_{ab}^{(1)} (N, \chi) =& \, \Sigma_{ab}^{(1;2)} (N) \ln^2(\chi) +
\Sigma_{ab}^{(1;1)} (N) \ln(\chi) , \\ 
\Sigma_{ab}^{(2)} (N, \chi) =& \, \Sigma_{ab}^{(2;4)} (N) \ln^4(\chi) + 
\Sigma_{ab}^{(2;3)} (N) \ln^3(\chi) + \nonumber \\ 
& \, \Sigma_{ab}^{(2;2)} (N) \ln^2(\chi) + \Sigma_{ab}^{(2;1)} (N)
\ln(\chi) .
\end{align} 
\label{eq:expansion} 
\end{subequations}
The functions $\Sigma^{(i;j)} (N)$ are purely functions of $N$. In order to match
\Eq{eq:truncated_expr} to the fixed-order results, we need to Fourier-invert
logarithms of the form $\ln^k (\chi)$, such computation requires the evaluation
of the following integral at a given order in $k$ 
\begin{equation}
\mathfrak{I}_k (N, \xi_p) =  \int\limits_0^\infty \mathrm{d} \hat{b}
\frac{\hat{b}}{2} J_0 \left( \hat{b} \sqrt{\xi_p} \right) \mathrm{ln}^k
\left( \bar{N}^2 + \frac{\hat{b}^2}{b_0^2} \right) .  
\end{equation} 
Thus, according to \Eq{eq:bhat-inverse}, the computation of $\mathfrak{I}_k$ for a
fixed value of $k$ just amounts to the computation of the following limit
\begin{equation} 
\mathfrak{I}_k (N, \xi_p) = \lim\limits_{\epsilon \to 0}
\frac{\partial^k}{\partial \epsilon^k} M(N, \xi_p, \epsilon) ,
\label{eq:inverse-fourier-lchi} 
\end{equation} 
with $M$ expressed as in \Eq{eq:bhat-inverse}. This allows us to analytically 
compute the inverse Fourier of the finite order expansion of the resummed expression 
up to $k = 4$,
\begingroup
\allowdisplaybreaks
\begin{subequations} 
\label{eq:finite-order-limit} 
\begin{align}
\mathfrak{I}_1 (N, \xi_p) =& -\frac{2 N}{\sqrt{\xi_{\mathrm{p}}}}
\mathrm{K}_{1}(2 N \sqrt{\xi \mathrm{p}}) \\ 
\mathfrak{I}_2 (N, \xi_p) =& -\frac{4 N}{\sqrt{\xi_{\mathrm{p}}}} 
\mathrm{K}_{1}^{(1)}(2 N \sqrt{\xi \mathrm{p}})-\frac{4 N}{\sqrt{
\xi_{\mathrm{p}}}}\left(\ln \left(\frac{N}{\sqrt{\xi_{\mathrm{p}}}}
\right )+\gamma_{E}\right) \mathrm{K}_{1}(2 N \sqrt{\xi_{\mathrm{p}}}) \\ 
\mathfrak{I}_3 (N, \xi_p) =& -\frac{6 N}{\sqrt{\xi_{\mathrm{p}}}} 
\mathrm{K}_{1}^{(2)}(2 N \sqrt{\xi_{\mathrm{P}}})-\frac{12 N}{\sqrt{
\xi_{\mathrm{p}}}}\left(\ln \left(\frac{N}{\sqrt{\xi_{\mathrm{p}}}}\right) 
+\gamma_{E}\right)\mathrm{K}_{1}^{(1)}(2 N \sqrt{\xi_{\mathrm{p}}}) \nonumber \\ 
& -\frac{6N}{\sqrt{\xi_{\mathrm{p}}}}\left( \ln^{2}\left(\frac{N}{\sqrt{
\xi_{\mathrm{p}}}}\right)+2 \gamma_{E} \ln\left(\frac{N}{\sqrt{\xi_{\mathrm{p}}}}
\right) - \zeta_2 + \gamma_{E}^{2}\right)\mathrm{K}_{1}(2 N \sqrt{\xi_{\mathrm{p}}}) \\ 
\mathfrak{I}_4 (N, \xi_p) =&-\frac{8 N}{\sqrt{\xi_{\mathrm{p}}}} \mathrm{K}_{1}^{(3)}
(2 N \sqrt{\xi \mathrm{p}})-\frac{24 N}{\sqrt{\xi_{\mathrm{p}}}}\left( \ln
^{2}\left(\frac{N}{\sqrt{\xi_{\mathrm{p}}}}\right)+2 \gamma_{E} \ln
\left(\frac{N}{\sqrt{\xi_{\mathrm{p}}}}\right)- \zeta_2 + \gamma_{E}^{2}\right)
\nonumber \\ & \mathrm{K}_{1}^{(1)}(2 N \sqrt{\xi \mathrm{p}})-\frac{24
N}{\sqrt{\xi_{\mathrm{p}}}}\left(\ln
\left(\frac{N}{\sqrt{\xi_{\mathrm{p}}}}\right)+\gamma_{E}\right)
\mathrm{K}_{1}^{(2)}(2 N \sqrt{\xi_{\mathrm{p}}}) -\frac{8
N}{\sqrt{\xi_{\mathrm{p}}}} \mathrm{K}_{1}(2 N \sqrt{\xi_{\mathrm{p}}})
\nonumber\\ & \left( \ln ^{3}\left(\frac{N}{\sqrt{\xi_{\mathrm{p}}}}\right)+3
\gamma_{E} \ln ^{2}\left(\frac{N}{\sqrt{\xi_{\mathrm{p}}}}\right)+ 3 \left(
\gamma_{E}^{2}- \zeta_2 \right) \ln
\left(\frac{N}{\sqrt{\xi_{\mathrm{p}}}}\right)- 3 \gamma_{E} \zeta_2 +
\gamma_{E}^{3} + \zeta_3 \right) , 
\end{align} 
\end{subequations} 
\endgroup
where the $\mathrm{K}_{1}^{(i)}$ are the $i$-th derivatives of the Bessel function
w.r.t. the argument. One can check that in the small-$p_T$ limit--meaning taking
$N \to \exp(- \gamma_E)$--the above expressions exactly reproduce the standard
small-$p_T$ expressions given by Eqs.~(121)-(124) of Ref.~\citep{Bozzi:2005wk}.
The equivalence between taking $p_T \to 0$ and setting $N \to \exp(- \gamma_E)$
deserves further comment. On can check that the modified logarithmic variable 
$\ln (1 + \hat{b}^2 / b_0^2)$ of Ref.~\citep{Bozzi:2005wk} is obtained from our 
modified $\ln (\chi (\hat{b}))$ by setting $\bar{N} = 1$ which in 
\Eqs{eq:finite-order-limit} translates into taking $N \to
\exp(- \gamma_E)$.

\section{Phenomenological studies} 
\label{sec:pheno}

In the previous sections, we have developed a framework that allows us to numerically 
perform the inverse Fourier-Mellin transform of the resummed expression and to match it to the
fixed-order result. We now turn to the main result of this paper, namely a detailed 
phenomenological study of the impact of the threshold-improved transverse momentum and 
combined resummation on transverse momentum distributions in the case of Higgs and Z-boson 
production. Our aim is to assess the potential impact of: (i) the improved transverse momentum
resummation through the inclusion of the soft contributions which we expect to
be more visible in the small-$p_T$ region, and (ii) the soft resummation at fixed-$p_T$
which we expect to improve the matching to fixed-order in the intermediate and 
large-$p_T$ regions.

All results are produced using the \texttt{NNPDF31$\_$nnlo$\_$as$\_$0118} set of
parton distributions from NNPDF3.1~\cite{Ball:2014uwa} through the 
LHAPDF~\cite{Buckley:2014ana} interface in the manner described in
Refs.~\cite{Bonvini:2010tp, Bonvini:2012sh} by expanding the parton luminosity
in the basis of the Chebyshev polynomials; we expect the results to be largely
independent of the PDF set used. In order to assess perturbative uncertainties and 
higher order corrections, we perform standard variations of factorization and
renormalization scales using the seven-point method. Notice, however, that the
present studies do not include resummation scale variation as in 
Ref.~\cite{Bozzi:2005wk}, which will be done in future studies at N$^3$LL.  In addition to 
defining a proper value to the cutoff $C$ entering in the Borel prescription, one also 
has to define the free parameters entering the matching function in \Eq{eq:matching_function}. 
The following results are produced by setting $k = 3$ and $m = 2$. There is some 
arbitrariness in the definition of $k$ and $m$; however, we did explicitly check 
that results do not change provided that $k$ is chosen from $2$ to $5$ and $m < k$.  
The resummed results are produced from our own resummation code while the
fixed-order part is provided by the NNLOJET~\cite{Cruz-Martinez:2018zbi}
implementation.

\subsection{Threshold-improved $p_T$ resummation in the large-$\hat{b}$ limit}
\label{subsec:standard-css}

\begin{figure}[!htbp]
\captionsetup[subfigure]{aboveskip=-1.5pt,belowskip=-1.5pt} 
\centering
\begin{subfigure}{0.485\linewidth}
\includegraphics[width=\linewidth]{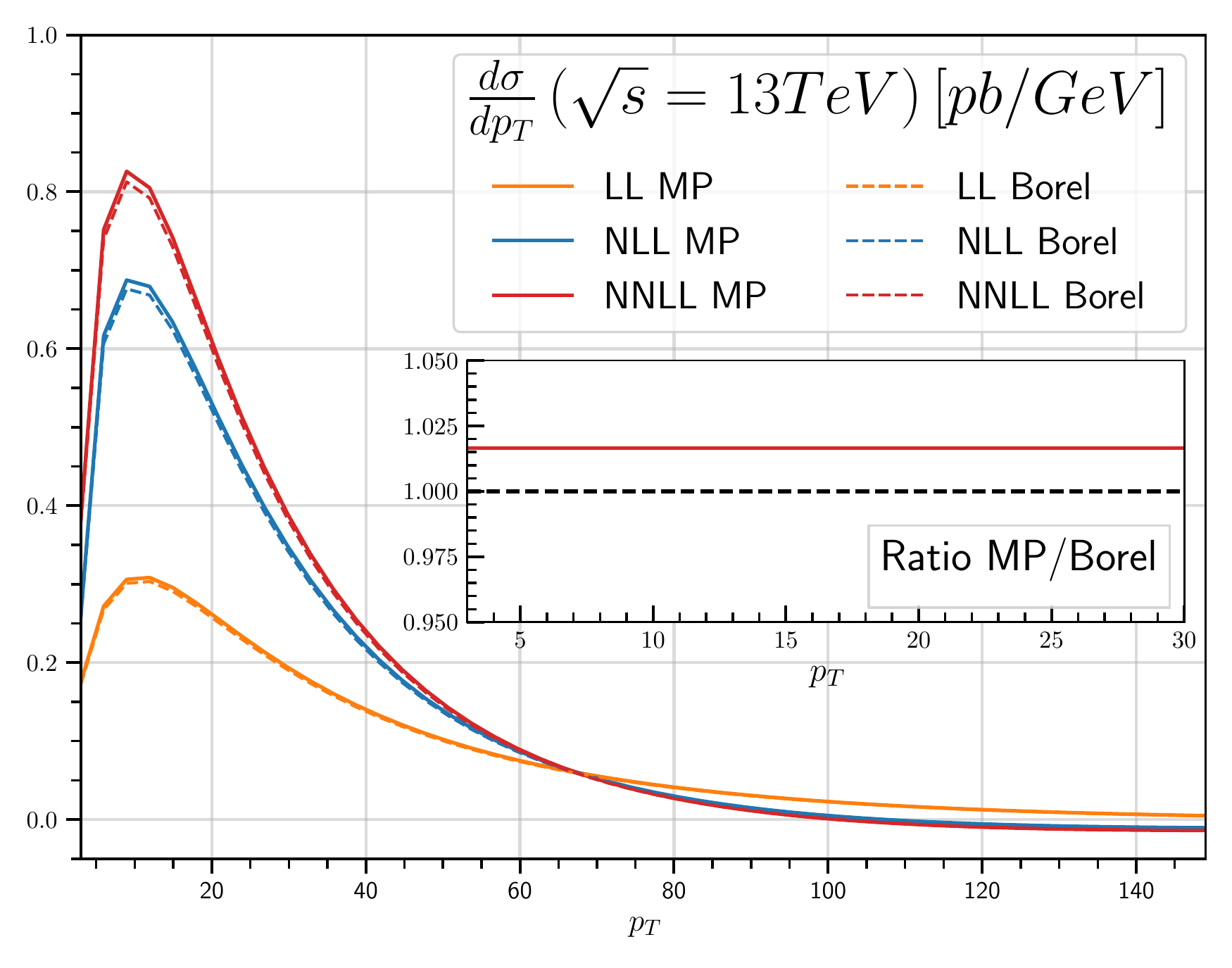} 
\end{subfigure} 
\hfil
\begin{subfigure}{0.485\linewidth}
\includegraphics[width=\linewidth]{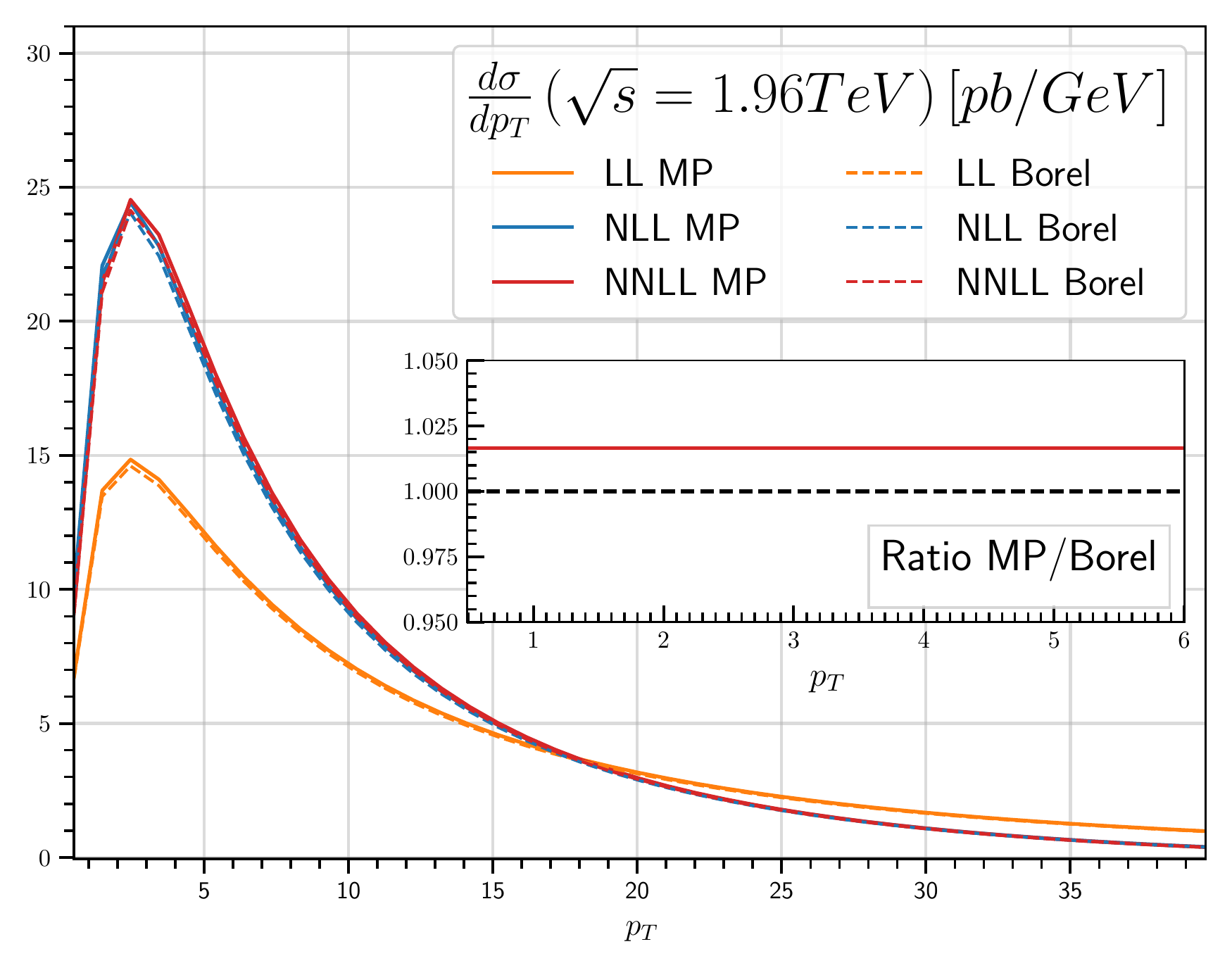} 
\end{subfigure}
\caption{Comparison of the Minimal and Borel prescription for the transverse 
momentum distribution of the Higgs (\textit{left}) and $Z$ boson (\textit{right}). 
The renormalization ($\mu_R$) and factorization ($\mu_F$) scales are set equal to 
the Higgs mass ($m_H$) and the $Z$ boson mass ($m_Z$) in the case of Higgs and DY 
production respectively. The normalizations are fixed to the total cross section.} 
\label{fig:min-vs-borel} 
\end{figure}

In this section, we start by checking that our results reproduce the standard
transverse momentum resummation when the soft contributions are switched off.
Besides being a consistency check, this allows us to verify that the Borel
prescription agrees with the Minimal Prescription (MP) of~Ref.~\cite{Bozzi:2005wk}.
It has been shown analytically in~Ref.~\cite{Muselli:2017bad} that threshold-improved 
$p_T$ resummation produces exactly standard $p_T$ resummation in the large-$\hat{b}$ limit, 
here, we show numerically that this is indeed the case using the Borel prescription. 
Taking $\hat{b} \to \infty$ means that we adopt the following replacements 
throughout our expressions: 
\begin{equation} 
\chi = \bar{N}^2 + \frac{\hat{b}^2}{b_0^2} \longrightarrow 1 + \frac{\hat{b}^2}{b_0^2}, 
\quad \text{and} \quad \frac{Q}{\chi} \longrightarrow \frac{b_0^2}{b^2}.  
\label{eq:large-b}
\end{equation}
Notice that in the expression of $\chi$, $\bar{N}$ has been replaced by $1$. As suggested 
in Ref.~\cite{Bozzi:2005wk}, shifting the argument of the logarithm by $1$ reduces the 
effect of unjustified small-$p_T$ resummation in the large-$p_T$ region (equivalently 
$\hat{b} \to 0$). The replacements in \Eq{eq:large-b} implies that we have to change the 
function $M$ in the following way, 
\begin{equation} 
\tilde{M}(\xi_p, \epsilon) = 2 \, \mathrm{e}^{2 \gamma_E \epsilon} \left( \frac{1}
{\sqrt{\xi_p}}\right)^{1+\epsilon} \frac{K_{1+\epsilon} \left( 2 \sqrt{\xi_p} \right)}
{\Gamma (-\epsilon)}.  
\label{eq:new-generating}
\end{equation} 
In order to check the consistency of the Borel formalism, we perform the replacement in \meqs{eq:large-b} 
and~(\ref{eq:new-generating}), and we implement the exact same procedure as in Ref.~\cite{Bozzi:2005wk}
by suitably adjusting the solutions to the evolution equation and fixing the integral over $p_T$
to be the fixed-order total cross section. The results shown in~\Fig{fig:min-vs-borel} demonstrates 
that our formalism is order by order consistent with the standard small-$p_T$ results from \texttt{HqT}~\cite{Bozzi:2005wk} for Higgs and \texttt{DYqT}~\cite{Bozzi:2008bb, Bozzi:2010xn} for 
DY which both use the Minimal Prescription. We note that apart from the changes mentioned above, no 
further treatment is needed for the Borel approach as the two formalisms only differ in the way the 
asymptotic nature of the resummed series is treated. This confirms two things: first, that the 
threshold-improved $p_T$ resummation indeed reproduces the standard small-$p_T$  resummation in the 
correct limit; and second, that the Borel method is a meaningful prescription to perform the inverse 
Fourier-Mellin transform.  

\subsection{Higgs boson production at LHC and DY production at Tevatron} 
\label{sec:higgs-pheno}

We begin by studying the standard model Higgs produced via gluon-gluon fusion at
LHC with a center of mass energy of $\sqrt{s}=\tev{13}$. The computations have
been performed in the large-top-mass ($m_T$) limit with a Higgs mass $m_H=\gev{125}$. 
As mentioned previously, the factorization and renormalization scales are 
varied according to the seven-point method by a factor of two in either direction 
around $m_H$.

\begin{figure}[!htbp]
\captionsetup[subfigure]{aboveskip=-1.5pt,belowskip=-1.5pt} 
\centering
\begin{subfigure}{0.475\linewidth}
\includegraphics[width=\linewidth]{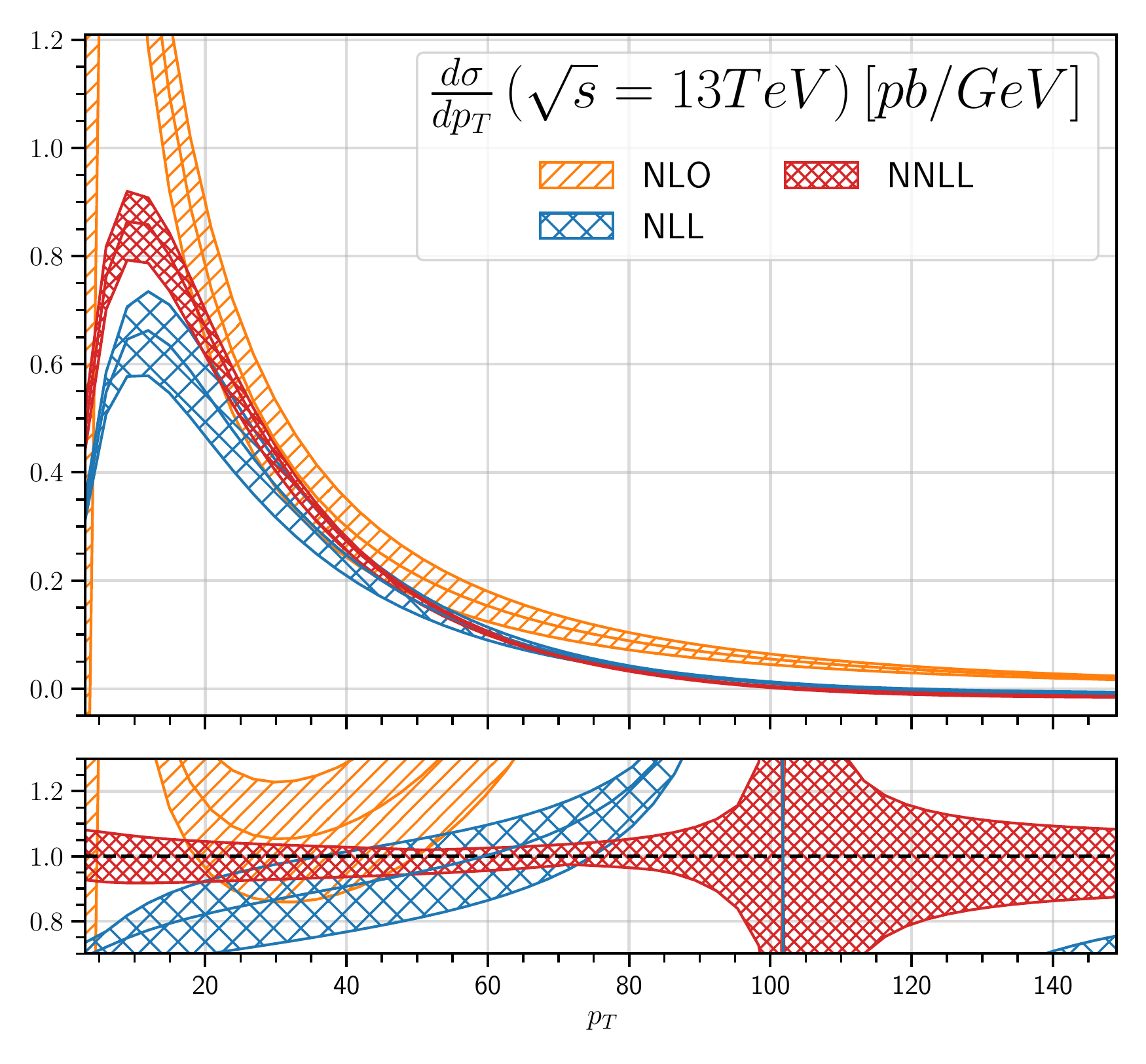}
\caption{SSpT resummation} 
\label{higgs:sspt} 
\end{subfigure} 
\hfil
\begin{subfigure}{0.475\linewidth}
\includegraphics[width=\linewidth]{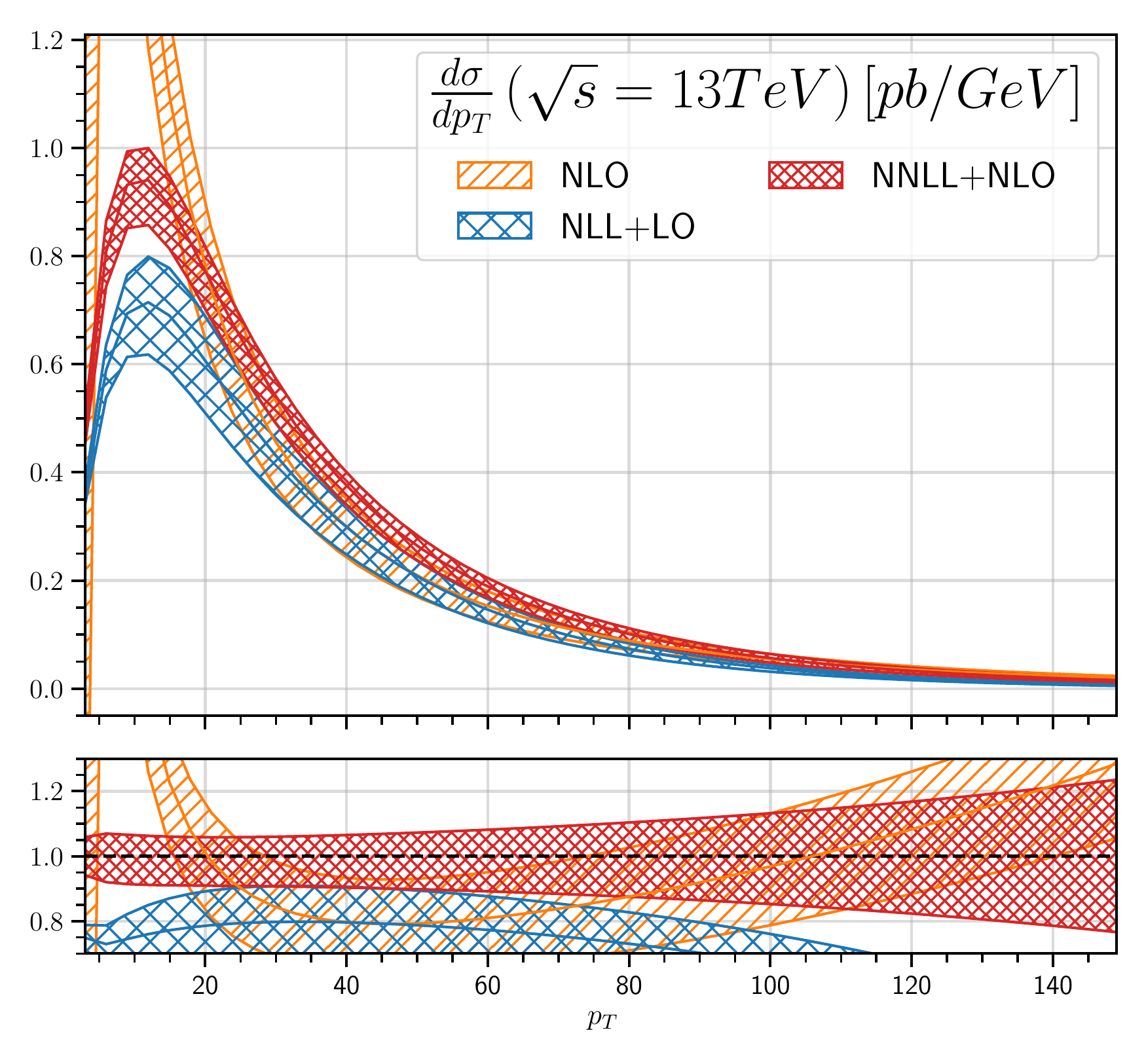}
\caption{SSpT resummation with matching} 
\label{higgs:sspt-matched}
\end{subfigure}
\begin{subfigure}{0.475\linewidth}
\includegraphics[width=\linewidth]{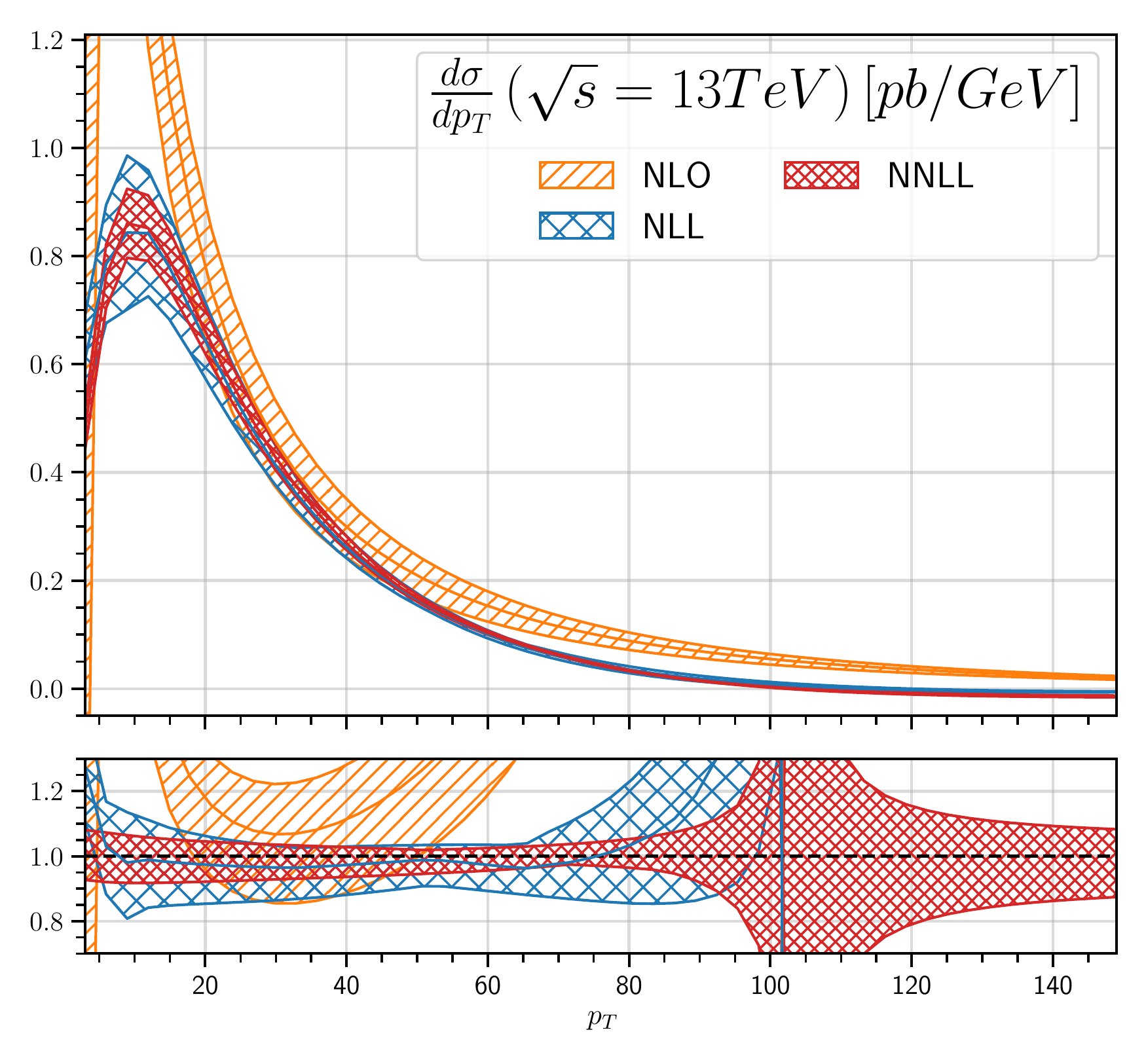}
\caption{TIpT resummation} \label{higgs:consistent} 
\end{subfigure}
\hfil \begin{subfigure}{0.475\linewidth}
\includegraphics[width=\linewidth]{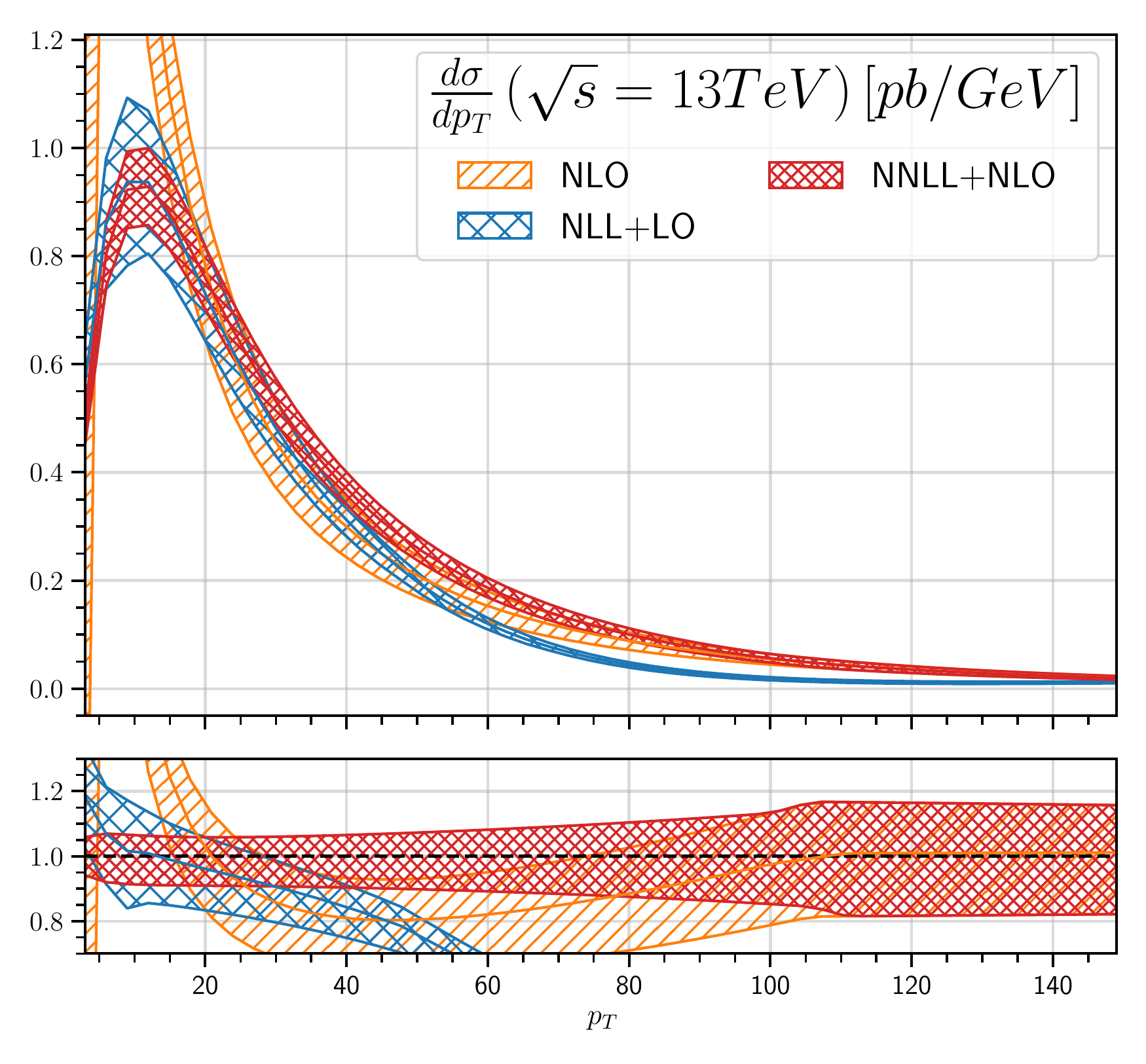}
\caption{TIpT resummation with matching}
\label{higgs:consistent-matched} 
\end{subfigure}
\begin{subfigure}{0.475\linewidth}
\includegraphics[width=\linewidth]{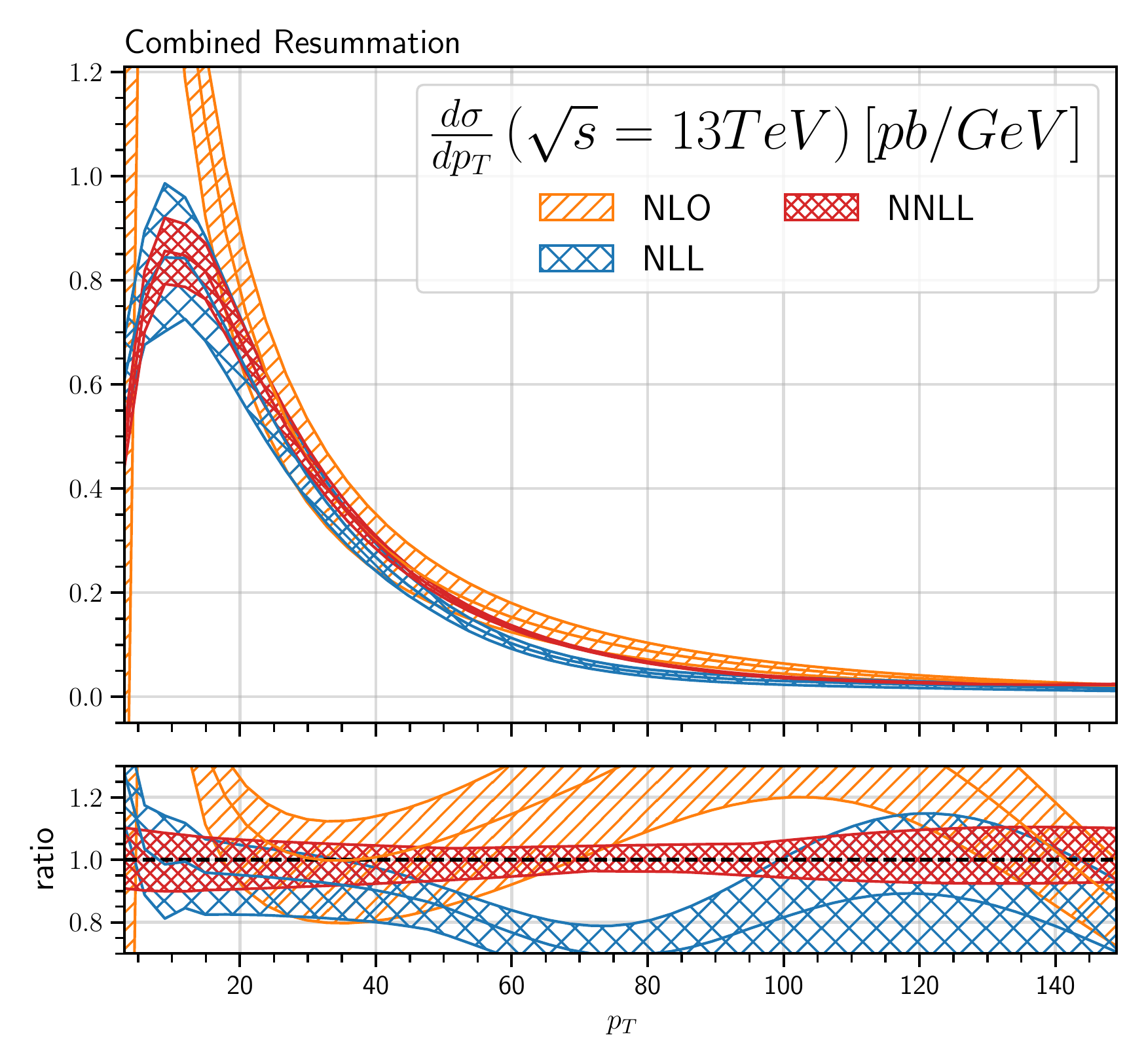}
\caption{combined resummation} 
\label{higgs:combined} 
\end{subfigure} 
\hfil
\begin{subfigure}{0.475\linewidth}
\includegraphics[width=\linewidth]{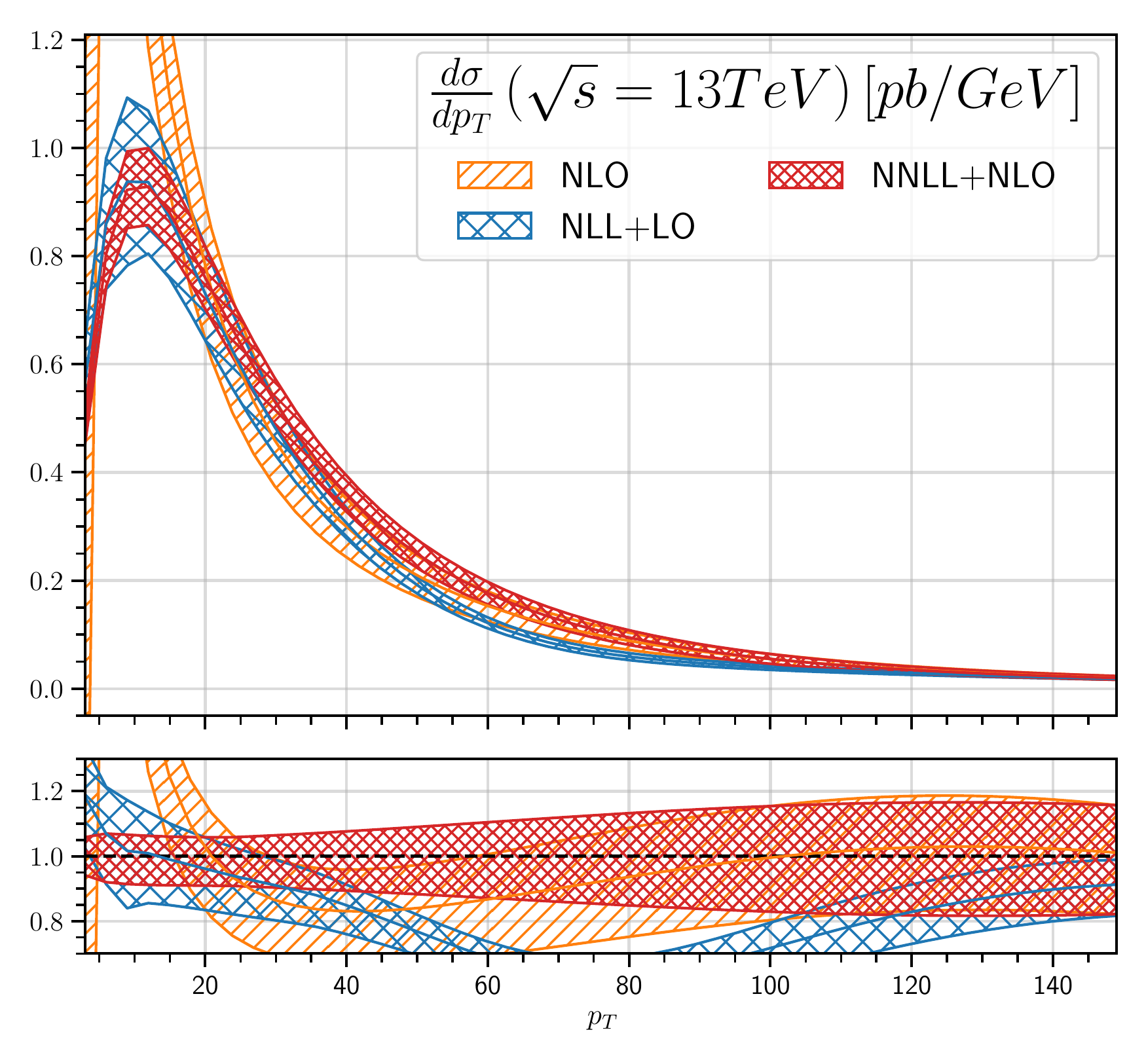}
\caption{combined resummation with matching} 
\label{higgs:combined-matched}
\end{subfigure}

\caption{$p_T$ spectrum of the Higgs boson production for various types of
resummation: standard small-$p_T$ (\textit{top}), threshold-improved small-$p_T$
(\textit{middle}), and combined (\textit{bottom}). The \textit{left} 
plots show the pure resummation results while those matched to the fixed-order 
are shown on the \textit{right}. The top panels compare the NLL and NNLL resummed 
results with the NLO. The lower panels show the ratio w.r.t. to the central value
of the respective NNLL results. The uncertainty bands are computed by varying 
$\mu_R$ and $\mu_F$ using the 7-point method; in all cases, $Q = m_H$.}
\label{fig:higgs-pheno} 
\end{figure}

The left plots of \Fig{fig:higgs-pheno} study the impact of the threshold improved
small-$p_T$ (TIpT, henceforth)
(\Fig{higgs:consistent}) and combined (\Fig{higgs:combined}) resummation on the
transverse momentum distribution by comparing the results to the NLO results. The
standard small-$p_T$ (SSpT, henceforth) resummation is shown in \Fig{higgs:sspt}. 
The top panels show the different order of resummation (NLL in blue and NNLL in red) 
along with the NLO results (orange). The ratio w.r.t. the NNLL result is shown in the
lower panels for the different types of resummation (SSpT, TIpT, and combined). 
In contrast to the NLO fixed-order results, the resummation leads to a well-behaved 
transverse momentum distribution that has peak at $p_T \sim \gev{10}$ and vanishes for 
$p_T \to 0$. As expected, we observe that the scale uncertainty is reduced 
with the inclusion of the NNLL terms. This feature is seen in all three types 
of resummations, but it is more pronounced in the small-$p_T$ region. Threshold-improved 
and standard $p_T$ resummation agree in the intermediate and large-$p_T$ region 
at NLL, but a sizeable difference appears in the small-$p_T$ range (below $\sim \gev{25}$).
Indeed, we see that the threshold-improved transverse momentum resummation displays faster
convergence at small values of $p_T$. As a consequence, in threshold-improved $p_T$ resummation,
the NLL result is much closer to the NNLL result than in SSpT resummation. However, at 
NNLL, the difference between SSpT and TIpT resummation is almost invisible, but with the 
threshold-improved $p_T$ resummation having a moderately smaller uncertainty. Turning now 
to the combined result, threshold-improved $p_T$ and combined resummation are exactly 
similar up to scales of at least \gev{40} where the pure threshold resummation starts to 
contribute. The inclusion of the threshold resummation leads to a surprising agreement 
between of the NNLL resummed result and the fixed-order, although a noticeable difference 
persists between the NLL and NLO results for larger values of $p_T$. Perhaps the most 
interesting feature of the combined resummation is the fact that even in the absence of 
matching the resummed computation seems to capture the behaviour of the fixed-order in 
the region where standard $p_T$ resummation fails to give accurate predictions. As a side 
note, unlike standard $p_T$ resummation, our modified resummation does not require any 
regulation procedure to get rid of its large-$p_T$ behaviour. Instead, it relies on the 
fact that its $\hat{b}=0$ limit coincides with the threshold resummed inclusive cross 
section that fixes the integral of the distribution.

Next, in order to fully obtain the most accurate predictions, we match the resummed 
expressions to the fixed-order results according to~\Eq{eq:matching-expression}. 
The results are presented on the right-hand side of~\Fig{fig:higgs-pheno}. The upper 
panels show the comparison between NLL+LO (blue), NNLL+NLO (red), and NLO (orange) for 
the three types of resummations. The lower panels show the ratio w.r.t. the
central scale (i.e. $\mu_R = \mu_F = m_H$) of NNLL+NLO. By 
comparing~\Fig{higgs:sspt-matched} with \Fig{higgs:consistent-matched}, we see faster 
convergence for the threshold-improved $p_T$ resummation. However, we notice that in the
absence of the additional threshold resummed expression, the NLL+LO threshold-improved $p_T$ 
resummation underestimates corrections from missing higher orders. This can be seen
in~\Fig{higgs:consistent-matched} where the NLL+LO band shifts away from both the
NLO and NNLL+NLO results. At small $p_T$, threshold-improved $p_T$ and combined 
resummation are exactly similar. The difference only occurs in the medium-range-$p_T$ 
where the contribution from the soft resummation again improves the agreement of the 
combined with the fixed-order result. In particular, we would like to highlight the good 
agreement between the resummed and the fixed-order results at large values of $p_T$. 

\begin{figure}[!htbp]
\captionsetup[subfigure]{aboveskip=-1.5pt,belowskip=-1.5pt} 
\centering
\begin{subfigure}{0.475\linewidth}
\includegraphics[width=\linewidth]{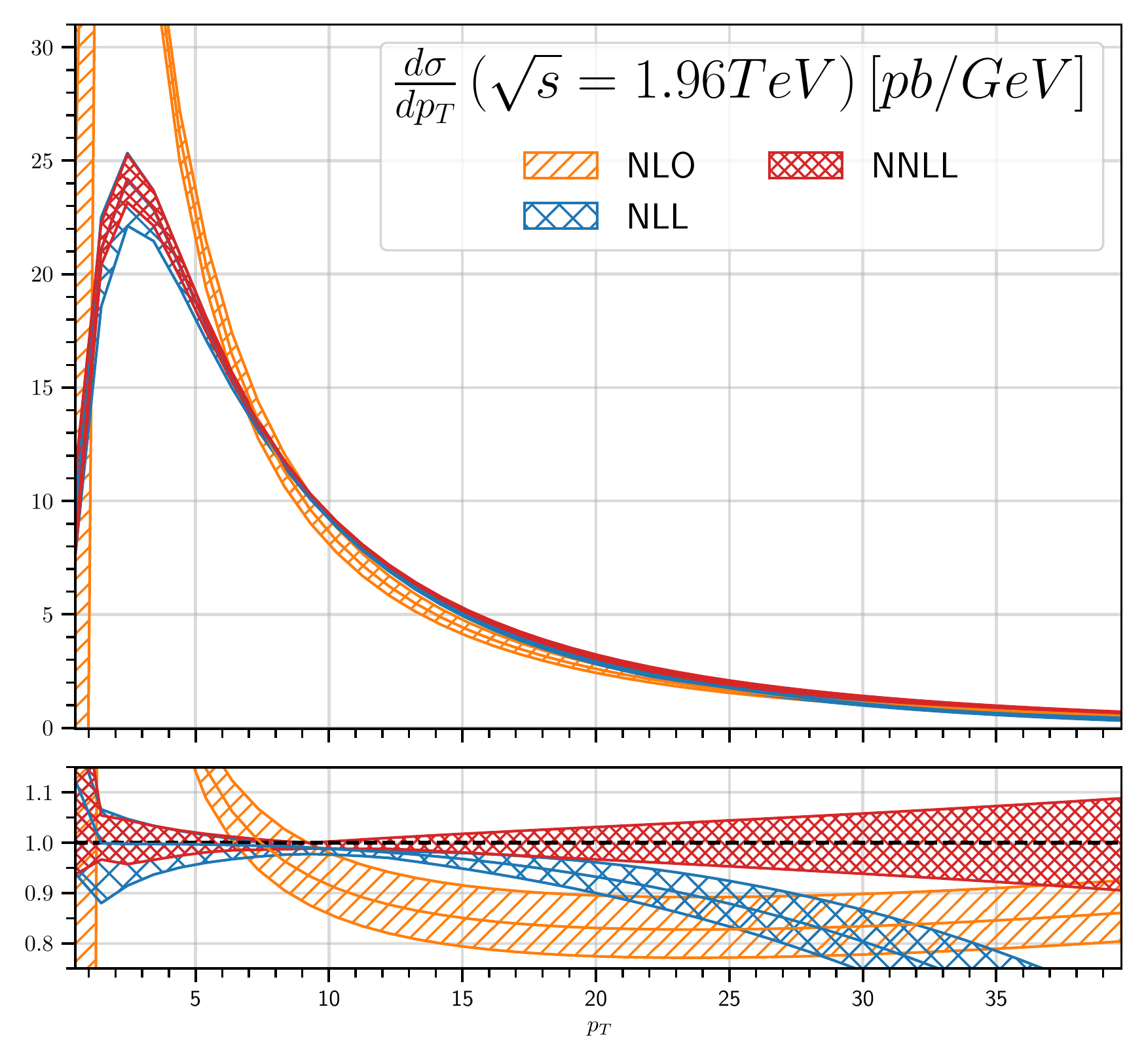}
\caption{SSpT resummation} 
\label{dy:sspt} 
\end{subfigure} 
\hfil
\begin{subfigure}{0.475\linewidth}
\includegraphics[width=\linewidth]{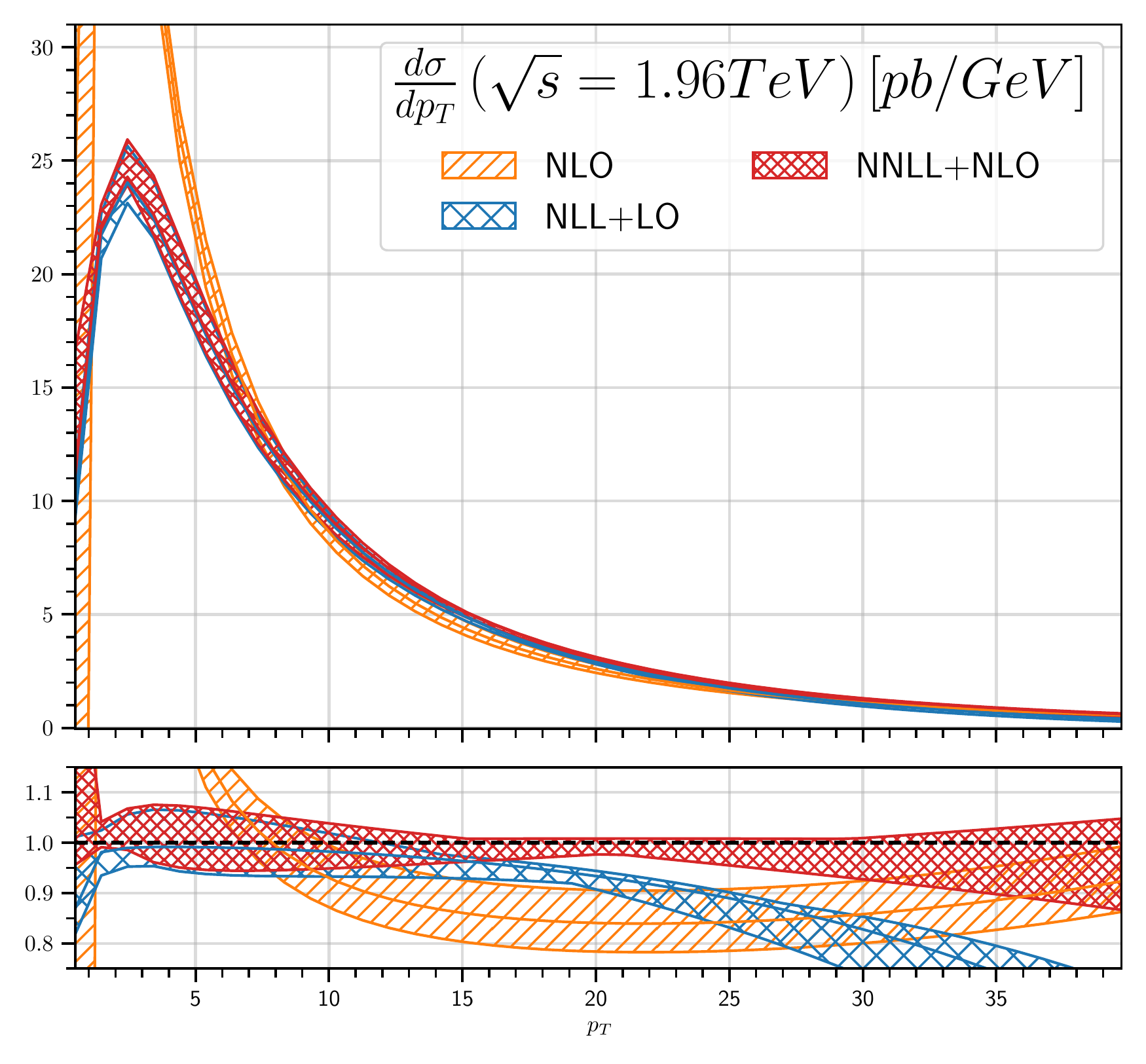}
\caption{SSpT resummation with matching} 
\label{dy:sspt-matched}
\end{subfigure}
\begin{subfigure}{0.475\linewidth}
\includegraphics[width=\linewidth]{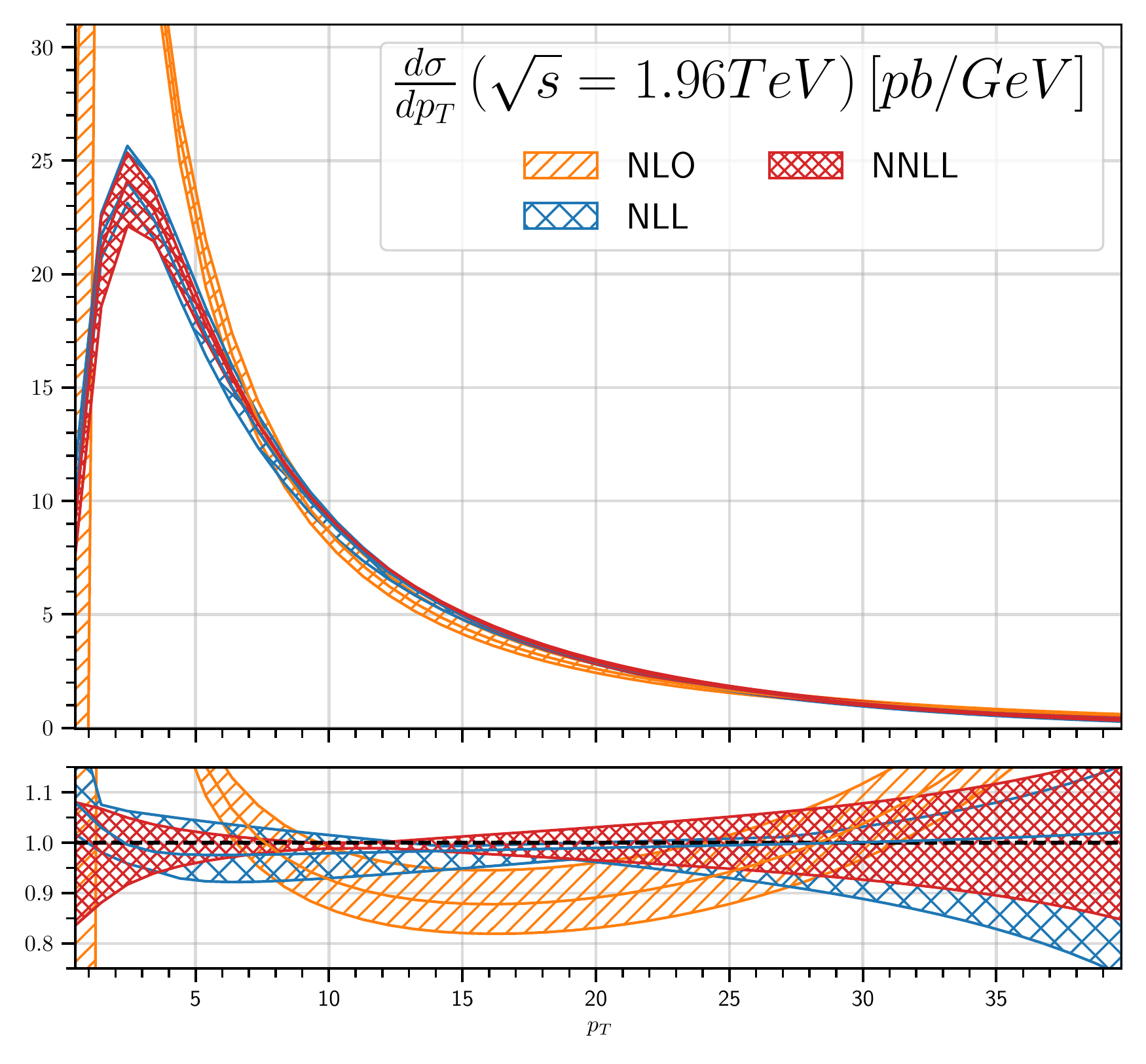}
\caption{TIpT resummation} \label{dy:consistent} 
\end{subfigure} 
\hfil
\begin{subfigure}{0.475\linewidth}
\includegraphics[width=\linewidth]{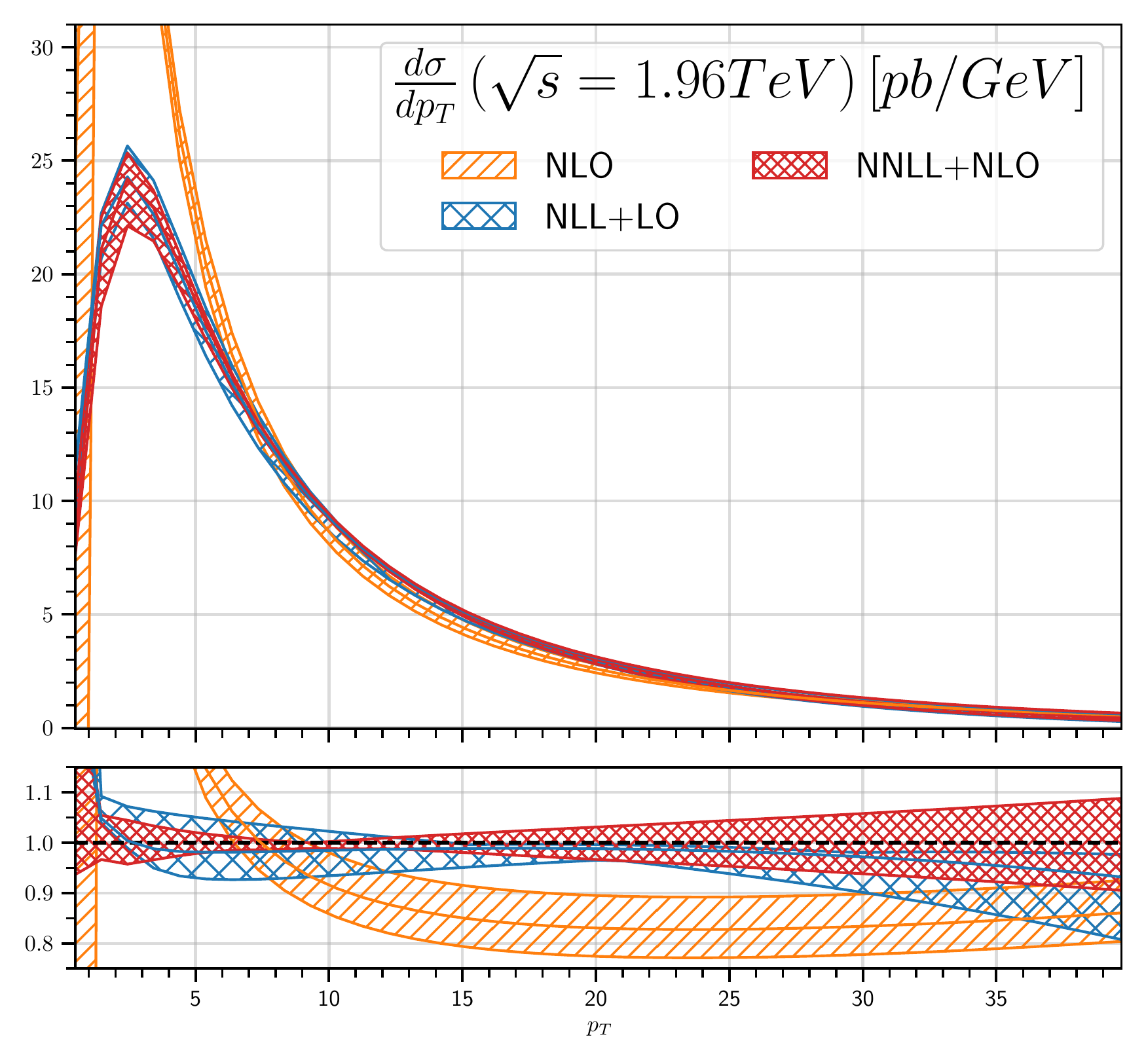}
\caption{TIpT resummation with matching} 
\label{dy:consistent-matched}
\end{subfigure}
\begin{subfigure}{0.475\linewidth}
\includegraphics[width=\linewidth]{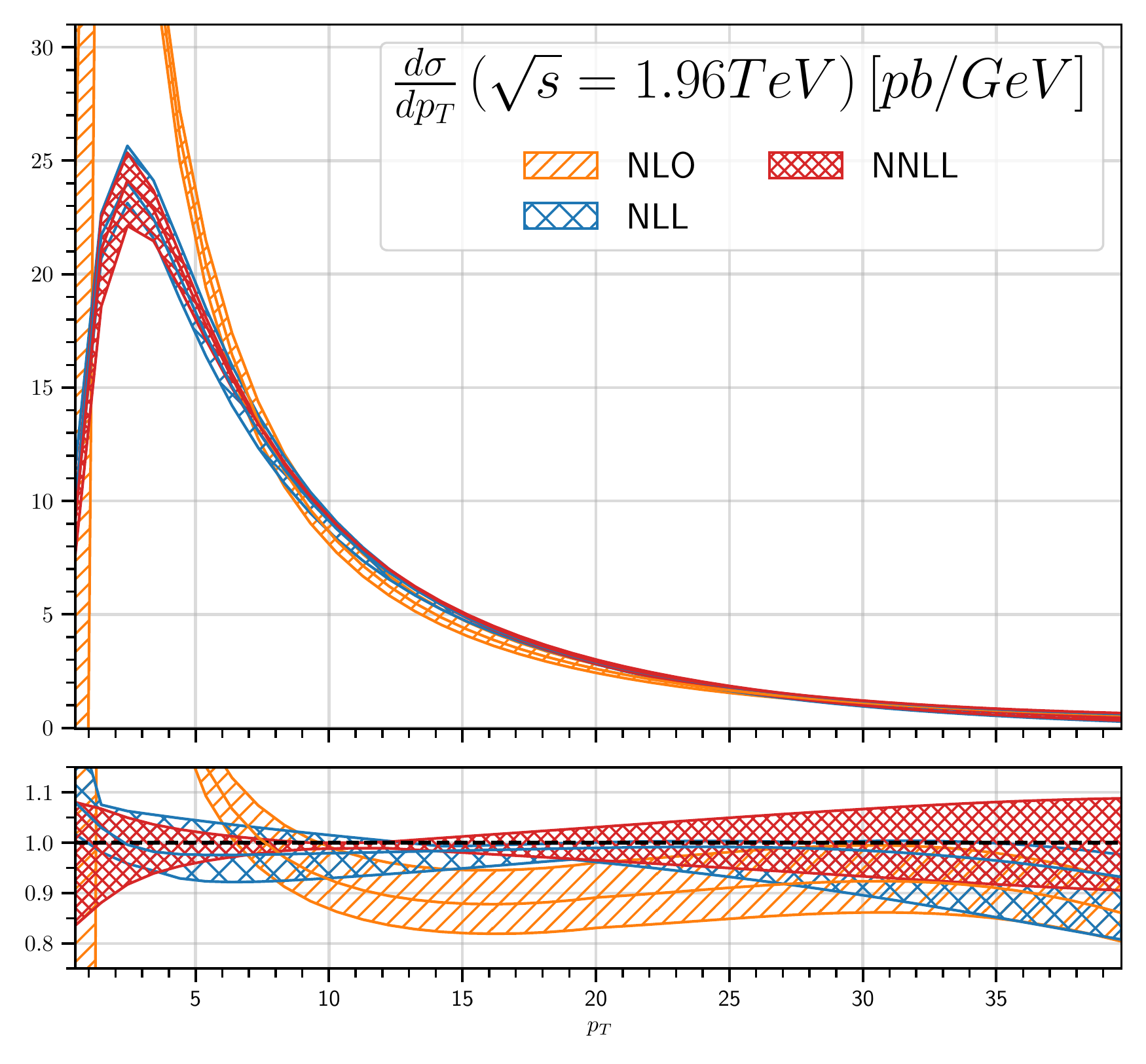}
\caption{combined resummation} \label{dy:combined} 
\end{subfigure} 
\hfil
\begin{subfigure}{0.475\linewidth}
\includegraphics[width=\linewidth]{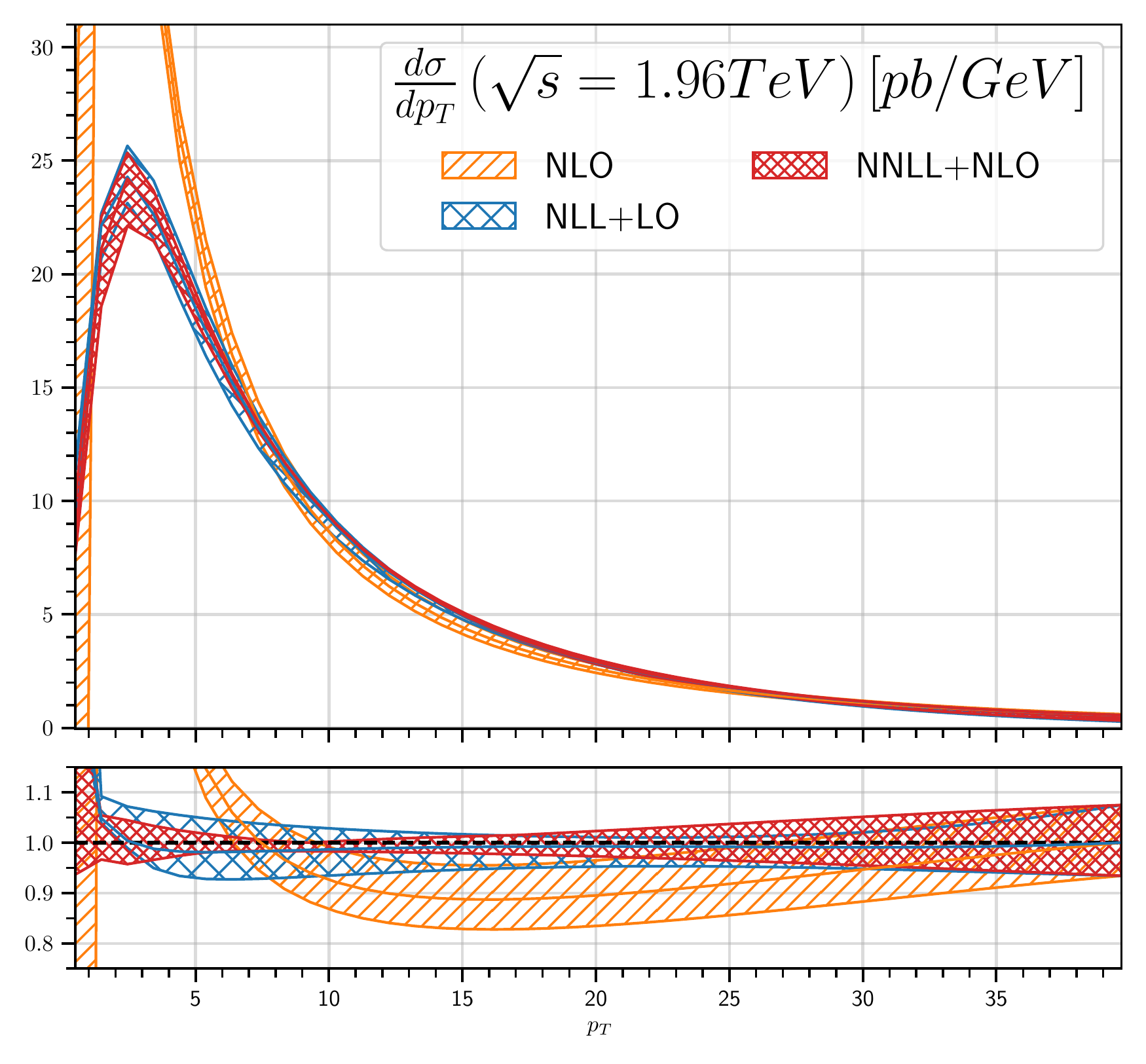}
\caption{combined resummation with matching} 
\label{dy:combined-matched}
\end{subfigure}

\caption{$p_T$ spectrum of the $Z$ boson production for various types of
resummation: standard small-$p_T$ (\textit{top}), threshold-improved small-$p_T$ 
(\textit{middle}), and combined (\textit{bottom}). The \textit{left} plots show 
the pure resummation results while those matched to the fixed-order are shown 
on the \textit{right}. The top panels compare the NLL and NNLL resummed results
to the NLO. The lower panels show the ratio w.r.t. to the central value
of the respective NNLL results. The uncertainty bands are computed by varying
$\mu_R$ and $\mu_F$ using the 7-point method; in all cases, $Q = m_Z$.}
\label{fig:dy-pheno} 
\end{figure}

In the context of threshold-improved $p_T$ and combined results, the matched results 
have broadly similar features to the purely resummed ones. The main new feature is that 
the matching of the combined resummation changes the result less than the matching of 
the TIpT resummation, and the latter less than the matching of the SSpT resummation.

We now turn to the application of the threshold-improved transverse momentum and 
combined resummation to a $Z$-boson production at the Tevatron via DY mechanism. 
We consider the production of a $Z$ boson with mass $m_Z = \gev{91.187}$ at the 
Tevatron with a center of mass energy $\sqrt{s} = \tev{1.96}$ because soft contributions 
for DY processes are expected to be more pronounced at colliders with lower center of 
mass energy.

\Fig{fig:dy-pheno} shows the results in comparison to the standard $p_T$
resummation. Similar to Higgs results in \Fig{fig:higgs-pheno}, the pure
resummed results are shown on the left-hand side. In all three types of
resummations, one sees similar features as previously, namely the fact that the
distributions go to zero as $p_T$ gets very small. The distributions then peak at
about $p_T \sim \gev{2}$ before vanishing again at large $p_T$. Both at NLL and
NNLL, threshold-improved $p_T$ resummation (\Fig{dy:consistent}) is analogous to 
standard transverse momentum (\Fig{dy:sspt}) at low $p_T$; small but yet noticeable
differences occur at large $p_T$ values, where the modified $p_T$ resummation displays
slightly better convergence. This feature is more accentuated in the combined
resummation (\Fig{dy:combined}) yielding a good agreement with the NLO (orange)
results. Again, these results suggest the validity of the threshold-improved 
$p_T$ and combined resummation even at large values of $p_T$ where the use of only
standard $p_T$ resummation is not fully justified.

This behaviour remains when we match the pure resummed calculation to the
fixed-order. The results are shown on the right-hand side of \Fig{fig:dy-pheno}.
In particular, we can notice the good agreement between the NLL+LO, NNLL+NLO,
and NLO results in the combined case (\Fig{dy:combined-matched}). Not only is 
the NNLL+NLO band smaller compared to the NLL+LO, but the latter is also
contained in the former suggesting a good convergence of the resummed
perturbative expansion. These results suggest that for the transverse momentum
distribution of the $Z$ boson where $x$ is often far from unity, the effect of
the threshold resummation is less pronounced.

\section{Conclusions} \label{sec:conclusion}

In this paper we have presented a phenomenological application of the combined
threshold and threshold-improved transverse momentum resummation introduced in 
Ref.~\cite{Muselli:2017bad} for transverse momentum spectra. This combined 
resummation allows for an improvement of the transverse momentum distribution 
that holds for all values of momenta.

We have presented in Section~\ref{sec:analytic-formulation} various analytical 
frameworks needed for the phenomenology studies. A particular attention was focused 
on the implementation of the Borel method as an alternative prescription to perform the 
inverse Fourier and Mellin transform. It was shown that the Borel prescription owes good 
numerical stability while providing complete control of the power correction terms, 
introduced by the threshold-improved $p_T$ resummation procedure, allowing us to reproduce 
standard $p_T$ results in the limit $p_T \to 0$. The results of our phenomenological studies 
were presented in Section~\ref{sec:pheno} where we matched our resummed expression with 
fixed-order calculations. We found that the effects of the threshold-improved $p_T$ resummation 
have a modest significance for a $Z$ boson production via DY mechanism at small and 
moderate $p_T$ while sizeable improvement can be seen in the tail of the distribution. 
The difference, however, is rather significant in the case of Higgs boson produced via 
gluon fusion, which highlights the potential relevance of threshold resummation to 
gluon-induced processes~\cite{Appell:1988ie, Giele:2002hx, Catani:2003zt, Ravindran:2003um}. 
As a main result we found that while threshold-improved $p_T$ resummation enhances the 
convergence at small-$p_T$, its combination with threshold resummation improves the agreement
with fixed-order results in the intermediate and large-$p_T$ regions.

As possible directions for future development of this work, we will extend the result
in Ref.~\cite{Muselli:2017bad} to N3LL and interface our implementation with Monte Carlo 
codes in order to produce N3LL+NNLO predictions. In addition, it will be interesting to study 
further the relation between threshold and the modified $p_T$ resummation in order to identify 
the missing soft logarithms with an ultimate goal of developing a combined expression that 
does not rely on a profile matching function. This, in turn, can be extended to account for 
the high-energy (or small-$x$) resummation, which quite recently has been jointly performed 
with small-$p_T$ \cite{Marzani:2015oyb} and threshold resummation \cite{Bonvini:2010tp}.

\acknowledgments{The author wish to thank Stefano Forte and Claudio Muselli for
useful discussion on combined resummation, and Giancarlo Ferrera for providing
help on standard $p_T$ resummation. We also express our deepest thanks to Juan
M. Cruz-Martinez for the fixed-order results, Stefano Forte and Christopher
Schwan for careful readings of the manuscript, and Luca Rottoli for several
discussions. This work is  supported by the European Research Council under the
European Union's Horizon 2020 research and innovation Programme (grant agreement
n.740006).}


\bibliographystyle{JHEP} 
\bibliography{refs.bib}

\end{document}